# QoS Provisioning Using Optimal Call Admission Control for Wireless Cellular Networks

by

Md. Asadur Rahman

A thesis submitted in partial fulfilment of the requirements for the degree of Master of Science in Electrical and Electronic Engineering

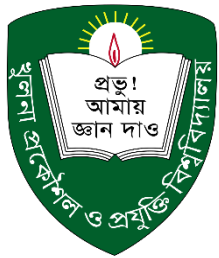

DEPARTMENT OF ELECTRICAL AND ELECTRONIC ENGINEERING
KHULNA UNIVERSITY OF ENGINEERING & TECHNOLOGY
KHULNA-9203, BANGLADESH
DECEMBER, 2014

# Declaration

This is to certify that the thesis work entitled "***QoS Provisioning Using Optimal Call Admission Control for Wireless Cellular Networks***" has been carried out by **Md. Asadur Rahman** in the Department of Electrical and Electronic Engineering, Khulna University of Engineering & Technology (KUET), Khulna, Bangladesh. The above thesis work or any part of this work has not been submitted anywhere for the award of any degree or diploma.

\-\-\-\-\-\-\-\-\-\-\-\-\-\-  	\-\-\-\-\-\-\-\-\-\-\-\-\-\-
Signature of Supervisor  	Signature of Candidate

# Approval

This is to certify that the thesis work submitted by **Md. Asadur Rahman** entitled "*QoS Provisioning Using Optimal Call Admission Control for Wireless Cellular Networks*" has been approved by the board of examiners for the partial fulfillment of the requirements for the degree of **Masters of Science (M. Sc.)** in the Department of **Electrical and Electronic Engineering**, Khulna University of Engineering & Technology (KUET), Khulna, Bangladesh in December 2014.

Thesis examination or oral examination committee

1. \_ \_ \_ \_ \_ \_ \_ \_ \_ \_ \_ \_ \_ \_ \_ \_ **Chairman**
   **Dr. Mostafa Zaman Chowdhury** **(Supervisor)**
   Assistant Professor
   Department of Electrical and Electronic Engineering
   Khulna University of Engineering & Technology (KUET)
   Khulna, Bangladesh.

2. \_ \_ \_ \_ \_ \_ \_ \_ \_ \_ \_ \_ **Member**
   **Head of the Department**
   Department of Electrical and Electronic Engineering
   Khulna University of Engineering & Technology (KUET)
   Khulna, Bangladesh.

3. \_ \_ \_ \_ \_ \_ \_ \_ \_ \_ \_ \_ \_ \_ \_ **Member**
   **Prof. Dr. Md. Nurunnabi Mollah**
   Department of Electrical and Electronic Engineering
   Khulna University of Engineering & Technology (KUET)
   Khulna, Bangladesh.

4. \_ \_ \_ \_ \_ \_ \_ \_ \_ \_ \_ \_ \_ \_ \_ \_ \_ \_ **Member**
   **Prof. Dr. Mohammad Shaifur Rahman**
   Department of Electrical and Electronic Engineering
   Khulna University of Engineering & Technology (KUET)
   Khulna, Bangladesh.

5. \_ \_ \_ \_ \_ \_ \_ \_ \_ \_ \_ \_ \_ **Member**
   **Prof. Dr. S. M. Abdur Razzak** **(External)**
   Department of Electrical and Electronic Engineering
   Rajshahi University of Engineering & Technology (RUET)
   Rajshahi, Bangladesh.

***Dedicated to My Parents***

Md. Anisur Rahman & Selina Rahman
Those who have chosen underprivileged life to continue my smile.

# Acknowledgements

*First of all, I would like to bend myself to Almighty for giving me the strength and confidence to complete my thesis work, successfully.*

*Then I would like to acknowledge my deep gratitude to my graduation committee.*

*This thesis work marks an important period in my life during which many people have played an important role. Although the author's name is the only one to appear on the cover, it is the combined effort of these people that helped me to complete this research work. Therefore, I would like to take this opportunity to thank several individuals for their contributions, realizing that such a list can never be complete.*

*It is a great pleasure to express my deepest gratitude and profound respect to my honourable supervisor,* **Dr. Mostafa Zaman Chowdhury***, Assistant Professor, Department of Electrical and Electronic Engineering, KUET, who fully gave many invaluable advices, proper guidance, constant encouragement, constructive suggestions, and kind co-operation throughout the entire progress of this research work. Especially, our meetings during the last couple of months were of great support. During all those meetings, he talked with such enthusiasm about my research, presentation, publications, etc. those made it always very interesting. I enjoyed all our meetings and above all want to thank him again for all the good advices.*

*I am also so much grateful to* **Prof. Dr. Md. Nurunnabi Mollah***, the honourable Dean, Faculty of Electrical and Electronic Engineering, KUET, as well as the member of this thesis examination committee, for his loving and cordial advices to prepare my thesis proposal and final thesis paper. It is also a great opportunity for me to show my gratitude and thanks to my one of the most favourite teachers,* **Prof. Dr. Mohammad Shaifur Rahman***, Department of Electrical and Electronic Engineering, KUET, and honourable member of this thesis examination committee, who encouraged me, a lot, to work about this topic, when I was so confused to choose my thesis issue.*

*I wish to complement to all teachers of the Department of Electrical and Electronic Engineering, KUET for their co-operations and relentless encouragements in various ways throughout this work.*

*Finally, it is also my duty to express gratefulness and appreciations to my very loving friends* **Md. Azizur Rahman (Shawon)** *and* **Farzana Rahman***, as well as obviously my family members for their unswerving inspirations to complete my work.*

<div align="right">

Md. Asadur Rahman
December, 2014.

</div>

# Abstract

The increasing demand for advanced services in wireless networks raises the problem for quality of service (QoS) provisioning with proper resource management. In this research, such a provisioning technique for wireless networks is performed by Call Admission Control (CAC). A new approach in CAC named by Uniform Fractional Band (UFB) is proposed in this work for the wireless networks for providing proper priority between new calls and handover calls. This UFB scheme is basically a new style of handover priority scheme. Handover priority is provided by two stages in this scheme which help the network to utilize more resources. The first priority stage is fractional priority and the second stage is integral priority. Fractional priority is provided by the uniform fractional acceptance factor that accepts new calls with the predefined acceptance ratio throughout the fractional priority stage (fractional band of channels). Integral priority is given to the handover calls by reserving some channels only for handover calls. In this work, it is shown that UFB scheme proofs itself as optimum call admission technique which is concerned about not only the QoS but also the proper channel utilization with respect to conventional guard channel and fractional channel schemes. In this thesis work, conventional fixed and fractional guard channel based CAC schemes are studied literally and presented in this paper in very easy mathematical method. In addition, the handover call rate estimation and its impact on QoS provisioning is discussed widely to attain the optimum QoS in proposed handover priority scheme. In multiple services providing wireless network, excessive call blocking of lower priority traffic is very often event at very high traffic rate which is a concerning issue for QoS provisioning. To attain such QoS provisioning for multiple services, another CAC scheme is proposed in this research work. This scheme is recognised by Uniform Band Thinning (UBT) scheme which is based on uniform thinning technique (UTT) and this is quite similar idea as UFB scheme. In this scheme, a set of channels experiences the fractionizing policy. This scheme reduces the call blocking probabilities (CBP) of lower priority traffic classes without notably increasing the CBP's of the higher priority traffic classes. The analytical functions of this scheme are deduced in general form which is useful to deduce for any number of traffic classes. In addition, numerical analysis of the proposed UBT scheme shows that the performances in terms of call blocking probability, overall call blocking probability, and channel utilization are improved and optimised compared to the conventional fixed guard channel scheme.

Keywords: *Call admission control, call dropping probability, call blocking probability, quality of services, thinning schemes, acceptance factor, uniform fractional band, uniform band thinning, channel utilization, and traffic class.*

# Contents

# List of Figures

# List of Abbreviation

| | |
|---|---|
| Base Station | BS |
| Call Admission Control | CAC |
| Call Blocking Probability | CBP |
| Call Dropping Probability | CDP |
| Fixed Guard Band | FGB |
| Fractional Guard Channel | FGC |
| Limited Fractional Channel | LFC |
| Mobile Station | MS |
| Non-Priority Scheme | NPS |
| Quality of Service | QoS |
| Uniform Band Thinning | UBT |
| Uniform Fractional Band | UFB |
| Uniform Fractional Channel | UFC |
| Uniform Thinning Technique | UTT |

# CHAPTER 1

Introduction

## Chapter Outlines

- ❖ Motivation
- ❖ Problem statements
- ❖ Challenges
- ❖ Contribution of this research
- ❖ Thesis outline

## 1.1 Motivation

Modern civilization has achieved unbelievable pace for the blessing of communication system. The revolutionary communication systems are making the whole world closer day by day. There exists a number of communicating ways in this modern era like telephone, fax, television, radio, email, mobile telephone, video conferencing, and many more. These communicating systems can be broadly classified into two categories— (*i*) wired and (*ii*) wireless. The cellular communication system is the latest member of wireless communication techniques but obviously the most effective communicating medium of the present generation where a number of communicating services converge effectively. These service facilities are generating enormous number of users with time. That is why, this system faces a huge traffic by reason of providing integrated services, such as the voice, data, and different types of multimedia.

The demand for multimedia services over the air is increasing drastically which leads to wide design consideration of wireless cellular network. These different types or classes of traffic are not equally important in the aspect of service priority. Traffic like security, healthcare, banking, handover calls (handover call initiates where an ongoing call moves from one macro-cell to another), etc. are more important where non-real-time calls like data, voice messaging, text messaging, etc. are comparatively less important [1]-[3]. So, different types of traffic are to divide into several classes. During the resource management, the important classes of traffic are prearranged higher priority and comparatively less important calls are considered as lower priority. Blocking a lower priority call is preferred over blocking a higher priority call for maintaining the quality of service (QoS).

There are two key objectives in cellular network design—(*i*) to maximize the resource utilization and (*ii*) to provide high QoS to users [4]. According to this statement along QoS management, it is also very important to utilize the radio resources from the service provider's point of view. Call admission control (CAC) is such a provisioning technique to maintain the QoS among the several traffic classes by limiting the number of call connections into networks that reduces the blocking probabilities of higher priority traffic classes and also optimizes the channel utilization [5], [6].

## 1.2 Problem statements

There are a number of CAC schemes proposed by various researchers based on different aspects of service management. These CAC schemes are either fundamental or conditional. The fundamental CAC schemes are applicable in any kind of wireless network but the conditional schemes are designed concerning special purposes.

There are some fundamental CAC schemes like new call bounding scheme [7], [8], non-priority scheme, cut-off priority scheme [5], [9], general form of cut-off priority scheme [10] etc. Due to provide priority to any special type of calls, it is necessary to reserve some channels dedicated for that calls. The reserved channels are not used for low priority calls. Since the radio resource is limited, this channel reservation decreases the proper channel utilization where non-priority scheme is able to utilize most of the radio resources but unable to ensure the satisfied level of QoS.

At peak hours, call arrival rate reaches very high. On the other hand, at off peak hour call arrival proportion goes down to very lower rate. Implementing a fixed guard band with a set of reserved channels for higher priority calls is the cause of unutilised resources at off peak hour. This problem demands a dynamic channel reservation technique. Again the dynamic channel reservation or dynamic channel allocation can be implemented on the basis of statistical calls arrival rate or random call arrival rate or traffic awareness.

Another concerning issue to design a CAC scheme is the proper estimation of handover call rate. Otherwise the QoS optimization concept may be underestimated or overestimated. On the basis of a proper statistical modelling, the handover call rate should be taken in account to analyse the CAC scheme otherwise the numerical analysis of that scheme may deceive us to realize the appropriate network performances.

Some CAC schemes [3], [11], [12] are designed considering adaptive bandwidth utilization. In bandwidth utilization idea, the scheme is designed to overcome the problem of resource utilization. Though these schemes ensure to utilize the maximum radio resources arises problem to provide proper QoS. These schemes cannot provide priorities to the special classes of traffic. Sometimes, traffic priority does not depend upon revenue only. Some traffic like health service or fire brigade or police help demands priority on the basis of humanity. In these benchmarks, to optimize the QoS, adaptive bandwidth related schemes are to avoid.

Else, CAC scheme is designed for multiple classes of services on the basis of their priorities [1] where the total arrival calls are classified in different traffic classes. Such a CAC scheme gives different level of priorities to different classes of traffic to provide suitable QoS. In this case, the traffic division into several classes is also an important challenge. This problem can be solved considering the "*maximum payee will get maximum priority*" technique. Nevertheless, this cannot be appropriate for all the moment, because of service importance, business tact, user quality and quantity, government ordinance etc. On the other hand, these realities may be hampering issues to attain maximum profit for the service provider companies. This is why, considering all of these conditions, a CAC scheme is always designed at optimized form, not maximum.

Another important issue to design a CAC scheme is its operating algorithm. Sometimes the algorithm of the scheme becomes very complex that should be avoided. Otherwise, the network with complex algorithm may take more times to take decision at multifaceted call admission. It is preferable that the admission controlling technique becomes stress-free.

CAC scheme based on adaptability of network is proposed in [3], [13] where resource is managed through the different calls on the basis of the networks adaptability functions. So, network adaptability is also a necessary step to design a CAC scheme.

On the basis of some novel CAC schemes, several authors proposed QoS optimization technique on the basis of traffic awareness [14], probabilistic call arrival rate [15], cost rate of holding time [16], mobility-awareness [17], [18] etc. These categories of CAC schemes are used in particular purposes. Generally, it can be said that all of these proposed schemes aim to reduce the call blocking probability of higher priority as well as improve the channel utilization for provisioning the optimum QoS of the networks for special purposes individually.

As a result, there arises scope to propose a CAC scheme so that it will be suitable to attain the desired QoS considering the previously mentioned important considerations such as reduced call blocking probability based on priority, optimum utilization of radio resources (bandwidth), reduced system cost, and easier algorithm in operation. Else, the CAC scheme contains an option to change its characteristic on the basis of network condition. The CAC scheme will also be capable to convert its scheme to another available scheme easily.

## 1.3 Challenges

Among the all existing CAC schemes, there are two major groups of mechanism. The first one is on the basis of new calls and handover calls. This type of CAC schemes are analysed in [4], [6], [12], [19], [21], [22]. The aspect of QoS depends upon these two types of calls where the handover calls get more priority over the new calls. The second groups are in case of multiple services oriented cellular network. There consider a number of service classes of different priorities in the network. In this case, about four or five classes are considered where the every type of services provided by the network is a subset of one of these service classes. This style of CAC schemes are proposed in [1], [5], [13]-[17], [20], [23] etc. These CAC schemes reserve channels for different classes in different manner.

In previous discussion, it is mentioned that some CAC schemes are novel upon which various special purpose dependent schemes are designed. According to the previous works, a new challenge is to accept that is optimum channel utilization with consciously compromise the QoS. Another challenge is to propose a mathematical model as call admission controlling scheme for multiservice network where any number of classes can be considered.

On the basis of these considerations, it is a challenge to propose new style of CAC schemes that can be able to optimize the QoS with proper channel utilization for aforementioned two types of CAC schemes. The schemes would be able to implement itself in special purposes too.

## 1.4 Contribution of this research

In this research work, basic CAC schemes are analysed and their limitations are identified. Thereafter, the concepts to overcome the limitations are explained and deliberated. Consequently, new style of fractional guard channel scheme are presented those are recognized as uniform fractional band (UFB) [24] to optimize the handover priority scheme and uniform band thinning (UBT) [25] scheme to reach challenges for multiclass traffic oriented networks. The basic idea of these two schemes is suggested to place a set of uniform fractional channels (uniform fractional band) between two consecutive bands where the forbidden service class may access with a predefined uniform fractional acceptance rate.

In this scheme, only a set of channels are fractionized uniformly. The difference of this scheme from the fractional channel scheme [10] is that the fractionizing rate of the band is uniform independent of channel occupancy. The proposed theories are classified for both handover priority and multiservice admission control schemes. In this paper, the mathematical expressions are presented and explained very straightforwardly.

Besides, these schemes are able to cope itself for special purpose usage like traffic awareness, dynamic channel allocation, probabilistic call arrival rate on the basis of uniform acceptance factor. This factor can be varied according to the design of the network to get the desired performances.

The related algorithms of rudimentary CAC schemes are discussed briefly. The performances of the proposed schemes are analysed numerically as well as the other conventional CAC schemes are also scrutinized by the same process. The performances of the proposed schemes are compared with the existing CAC schemes graphically and elucidated the benefits as well as limitations of the proposed scheme.

## 1.5    Thesis outline

- **Chapter 2** contains the related works of this thesis and their limitations.
- **Chapter 3** comprehends the proposed handover priority CAC scheme and its performance comparisons with the other popular existing CAC schemes in the aspect of optimum QoS provisioning.
- **Chapter 4** describes the proposed uniform band thinning CAC scheme for multiservice wireless networks and the optimization process of its QoS.
- **Chapter 5** concludes the total work with clear explanation of applicability, benefits, and limitations of the proposed schemes. The scopes of future work are also mentioned in this chapter.

# CHAPTER 2

Contextual works

## Chapter Outlines

- Introduction
- Guard band based CAC schemes
- Fractional channel schemes
- Special purpose schemes
- Summary

## 2.1 Introduction

There are several kinds of CAC schemes proposed by different authors. Broadly these schemes can be divided into three categories. They are fixed guard band based schemes, fractional guard channel schemes and special purpose schemes. In this chapter, the concepts of these schemes are discussed shortly. These schemes are also divided into some subclasses. These subclasses are also discussed. Their operational facilities and limitations are also argued.

## 2.2 Guard band based CAC schemes

A number of guard band or guard channel based CAC schemes are available where the main concept is to reserve channels for handover calls. There are three different such schemes those are defined below in brief.

*i*) *Cut-off priority scheme*: The cut-off priority scheme [9], [26] blocks a new call if the number of free channels is less than the number of guard band scheme reserved for handover calls. It is very easy technique to attain the desired QoS. This scheme is not concerned about proper channel or bandwidth utilization [27], [28].

*ii*) *Rigid division based scheme*: The rigid division based scheme [29] divides all channels available in a cell into two groups: one for common use and the other exclusively for handover calls. This scheme is not concerned about proper channel utilization.

*iii*) *New call bounding scheme*: The new call bounding scheme [22] limits the number of new calls admitted to the cell. It is a dynamic decision admitting process on the basis of call arrival rate. To attain the desired QoS this scheme blocks more new calls than cut-off priority scheme. To generalise this idea for multiple services is a complex task.

All of the above schemes deal only with homogeneous traffic (systems in which each type of traffic has the same bandwidth requirements and identically distributed channel holding time with the same average value). The aforementioned CAC schemes are unavailable for wireless networks supporting multimedia services with diverse QoS constraints [30]. These fixed guard band based CAC schemes cannot assure the proper channel utilization due to its channel reservation for handover calls. Such CAC scheme is not applicable in high traffic multimedia services oriented wireless network due to these aforementioned obstacles.

## 2.3   Fractional guard channel schemes

This type of scheme admits call with certain probability on the basis of channel occupancy or arrival call rate or predefined acceptance factor. In the literature of fractional guard channel (FGC) scheme there are three types of fractional channel scheme. Basically, FGC scheme fractionizes the channels for a particular traffic class like new call that provides the priority to handover calls. Sometimes, the schemes are also known as thinning schemes [19], [22]. The fractional channel schemes are discussed below.

*i*) *Thinning scheme I*:  Thinning scheme I admits a new call with certain probability which depends on the number of busy channels. This scheme was first proposed in [10] and shown to be more general than the cut-off priority scheme. A moderate version of thinning scheme is limited fractional channel (LFC) scheme which is better than thinning scheme I [31]. The idea of this scheme cannot be applicable for multiple services network.

*ii*) *Thinning scheme II*:  Thinning scheme II admits a new call with certain probability based on the number of new calls accepted into the cell. This scheme was first proposed in [19] and shown to be more general than the new call bounding scheme. Thinning scheme II is not concerned about channel utilization.

*iii*) *Uniform fractional channel scheme*:  Uniform fractional channel (UFC) scheme is another type of thinning scheme. UFC scheme admits call with uniform acceptance factor regardless the channel occupancy or call arrival rate. This scheme is proposed in [32]-[34]. UFC scheme is only applicable in the network where the ratio of handover call and new call is very small [33].

Guard channel scheme or thinning scheme I reserves a number of resources (bandwidth/number of channels/transmission power/codes) for the exclusive use of handover attempts in cellular networks (this reduces the handover call dropping probability). However, due to the fact that an integer number of channels or bandwidth (basic units according to the network) are reserved in the conventional guard channel schemes, the blocking probabilities of the call types involved (i.e., new and handover calls) vary greatly as the number of reserved channels are changed. To overcome this problem, two schemes have been proposed: the LFC scheme [10], [31] and the UFC scheme [32]-[34]. Though the limitation of thinning scheme I is solved by LFC scheme but in this case, the authors in [10] do not discuss its channel utilization performance. The channel utilization problem can be solved by UFC scheme in higher traffic condition but it becomes the threat for QoS because UFC scheme can be applicable only in low handover and new call ratio oriented network [33].

Though thinning scheme is broadly discussed in [22] on the basis of handover call dropping probability and new call blocking probability, the optimized channel utilization is not confirmed by the authors. Else, thinning scheme I and II are generalized for multiservice network mathematically in [19] but its performances are not compared with the other fundamental CAC schemes.

## 2.3    Special purpose schemes

On the basis of novel CAC schemes, some authors proposed QoS optimization technique on the basis of some special purposes. In [14] the authors optimise the QoS of the network on the basis of traffic awareness but in this scheme fixed guard band scheme is used as a novel CAC scheme. On the concept of probabilistic call arrival rate, a CAC scheme is proposed in [15] where as a novel CAC scheme, thinning scheme II is used. Else, on the basis of cost rate of holding time in [16] and mobility-awareness dependent CAC schemes [18], [21] are not fundamental CAC schemes. These types of CAC schemes are used in particular purposes for specific networks but not in general cases.

## 2.4    Summary

Detailed study of the different types CAC schemes reveals the following facts:

- ➢ Guard band based schemes reduce the channel utilization.
- ➢ Thinning schemes are not concerned about channel utilization.
- ➢ UFC scheme cannot assure the QoS in high traffic.
- ➢ There exist a common complexity to approach a concept for two traffic and multiple traffic CAC scheme.
- ➢ Special purposes CAC schemes are not novel schemes.

Therefore, it is a burning challenge to a new of CAC scheme that can optimise the QoS with proper channel utilization and the scheme will be capable of network adaptability by its special characteristic that can help to convert itself into another form of CAC scheme.

# CHAPTER 3

Handover priority by UFB scheme

Chapter Outlines

- ❖ Introduction
- ❖ Hypothesis on the rate of handover call
- ❖ Fixed and Fractional guard channel schemes
- ❖ Proposed handover priority scheme
- ❖ Performance analysis
- ❖ Summary

## 3.1 Introduction

The cellular communication is one of the favourite ways for worldwide communication. The number of users of cellular system throughout the world is increasing drastically day by day. Due to incremental demand of the cellular networks, there has been tremendous interest and progress in this field. Cellular network provides services by dividing its physical area into different specific regions called cells. When a mobile user crosses the cell boundary or the quality of the wireless link between mobile station (MS) and base station (BS) is unacceptable, then the process of handover call is initiated [35]. In recent years, a remarkable tendency to design the cellular network is (*i*) decreasing the cell size and (*ii*) increasing the user mobility [3]. These two factors result in more frequent handovers. In practice, it is observed that users are more sensitive with a view to dropping of an ongoing and handed over call than blocking a new call [36]. A proper management of channel allocation is necessary for this kind of provisional service.

A CAC scheme is such a provisional technique to provide the QoS to the different calls at the target level by limiting the number of enduring calls in that system [36]. One major challenge in designing a CAC arises to provide service two major types of calls: new or originating calls and handover calls. The QoS performances related to these two types of calls are generally measured by new call blocking probability and handover call dropping probability. Since blocking a new call is less serious than dropping a handover call, CAC schemes usually give a higher priority to handover calls. Besides, optimum resource utilization (channel utilization) is also a very important issue to design a CAC because the resource of a wireless system is limited.

There are a number of CAC schemes [1], [2], [19], [22], [35]-[39] where the authors consider the different parameter to analyse the performance of their proposed schemes. A very general and easiest CAC scheme is new call bounding scheme and this is based on FGB scheme. By this method, QoS can be achieved easily but the scheme cannot assure the proper channel utilization for reserving some channels only for handover calls. To improve the QoS and to get other important system facilities several guard band schemes are proposed. FGC scheme or thinning scheme-I and LFC scheme are proposed in [10]. In FGC scheme, new calls are accepted by the channels from starting state to end depending on the channel occupancy and the acceptance factors vary at degrading manner from 1 to 0. New call blocking probability (CBP) and handover call dropping probability (CDP) increases excessively with the increasing call arrival rate in FGC scheme. Due to this characteristic, its channel utilization profile is very poor. In [10] the authors proposed LFC scheme for optimum QoS and with respect to fixed guard band (FGB) and FGC scheme but in this case channel utilization is not considered. Thinning scheme-II is proposed in [19], [22]. According to the thinning scheme-II new calls are accepted by different channels depending on the traffic arrival rate. By this scheme any level of QoS can be achieved with compromising the CBP and so why it cannot assure the optimum channel utilization. UFC scheme is proposed in [32]-[34]. Authors of these papers conclude that the UFC scheme is better in the lower traffic than FGB scheme but at the higher traffic its performance is very poor. On the other hand, the channel utilization factor of the UFC scheme is more than FGB, LFC, and FGC schemes.

In this chapter, a new style of guard band CAC scheme is proposed which is basically a hybridization of non-priority scheme (NPS) [7], UFC scheme [32]-[34], and priority scheme. This scheme is called in this work as UFB scheme. In this scheme, the total channels are divided into three bands. In first band, the new calls and handover calls get access with uniform acceptance factor, 1 (Acceptance factor is the measure of the call accepting ratio from the total arrival rate at a moment) and this band will show the non-priority characteristics. In second band, the new calls will be accepted by the existing channels with a predefined acceptance factor less than 1 which is the condition of fractionizing the band such a value may be any number within 0 to 1 and the handover calls will be accepted with unity acceptance factor. At last, the new calls will be accepted at the rate of void acceptance factor and the whole band channels accept only handover calls. In this scheme, handover calls get priority by two bands where in FGB scheme, there is only one band where handover calls get priority. For this reason, UFB scheme can accept some more new calls than the FGB scheme because there is a uniform fractional band inside the priority and non-priority bands. In this case, there is a trepidation fact to increase the CDP for decreasing the CBP. To overcome this obstacle, it is necessary to determine the proper value of acceptance factor of the middle band of this scheme. Moreover, UFB scheme ensures the more channel utilization than FGC, FGB, and LFC schemes.

Besides, the performances of the CAC schemes in [10], [19], [22], [32]-[34], [40] are analysed considering the handover calls as the fixed ratio of the new calls. In practice, the handover calls rate in a system cannot maintain a fixed ratio with new calls which is proposed in [3], [9]. Essentially, handover call estimation can be calculated by the equation of statistical hypothesis on handover calls and new calls which is explained at the next section in this chapter. The difference between the fixed ratio handover call and the statistical modelling of call handover rate is also mentioned by the numerical results at the performance investigation section.

Because of using a uniform fractional band between the priority and non-priority band, some more new calls will get access which must reduce the CBP but it may increase the CDP than FGC due to extra traffic. To avoid such a problem, necessary steps are discussed with proper explanations. This scheme uses a short uniform band which can avoid the excessive computational complexity because the uniform acceptance factor does not depend on the channel occupancy or call arrival rate through this band. So, it is very easy to observe the impact of various acceptance factors on CBP and CDP. The novelty of the acceptance factor is to determine the lower CBP for lower and higher call arrival rate that means this scheme can solve the limitation of UFC scheme. Furthermore, a comparison is made between the proposed CAC scheme and other different CAC schemes on the basis of CBP, CDP, channel utilization, and overall call blocking probability. It is also explained that why and whether the performance of the proposed UFB scheme is applicable and beneficiary in the aspect of optimum QoS.

## 3.2  Hypothesis on the rate of handover call

In cellular networks, the rate of new call and handover call does not maintain the fixed ratio. This is why a hypothesis is necessary to obtain the relation between them. The relation among the originating or new call arrival rate ($\lambda_n$), the handover call arrival rate ($\lambda_h$), and the average channel departure rate ($\mu$) is essential to determine the probability of blocking new calls and dropping handover calls. Here, it is considered that $P_B$ and $P_D$ represent the blocking probability of new calls and the dropping probability of handover calls, respectively. All calls arriving processes are assumed to be as Poisson's distributed.

A new call that arrives in the system may be either completed within the original cell or handed over to another cell before completing the call. The probability of a handover call depends on two factors - (*i*) the average dwell time (*1/η*) and (*ii*) the average call duration (*1/μ$_a$*). Again the average channel departure rate ($\mu$) also depends on the above two parameters. Since both the call duration and the cell dwell time are assumed to be exponential [3], [9], the handover probability, $P_h$ of a call at a particular time is given by,

$$P_h = \frac{\eta}{\eta + \mu_a} \tag{3.1}$$

Furthermore, the arrival rate of handover calls into a cell is evaluated as,

$$\lambda_h = \frac{(1-P_B)P_h}{[1-P_h(1-P_D)]}\lambda_n \tag{3.2}$$

The equation agrees from balancing the rates of handover calls into and out of a cell. When a call is originated in a cell, the call holds the channel until that call is completed in the cell or the mobile moves out of the cell. Therefore, the channel holding time, $T_c$ is either dwell time, $T_h$ or the call length time, $T_n$ [9]. The minimum value between dwell time and call length time is the channel holding time. Therefore, the relation among $T_c$, $T_h$, and $T_n$ can be represented as,

$$T_c = \min(T_h, T_n) \tag{3.3}$$

## 3.3  Fixed and Fractional guard channel schemes

There exist a number of CAC schemes based on the notion of guard channel where reservation of total channel between new calls and handover calls are assigned in various manners. In every case, handover calls get priority over new calls. At first the fundamental idea of fixed guard band CAC scheme is analysed. Thereafter, the fractional guard channel schemes are also described on the basis of one dimensional Markov chain. These studies are presented in easiest mathematical way as well as their admission controlling algorithms are also discussed gradually with neat state transition rate diagram.

### 3.3.1  Fixed guard band scheme

Fixed guard band scheme is a general priority scheme for call admission. In this case, priority is given to handover calls by assigning guard channels ($G_C$) entirely for handover calls only among the total channels in a cell say, $C$. The rest $M (= C - G_C)$ channels are shared by both new calls and handover calls. A new call is blocked if the progressive call is in state, $M$ or more than that. A handover call is blocked if no channel is accessible in the target cell that means the operating state is at $C$.

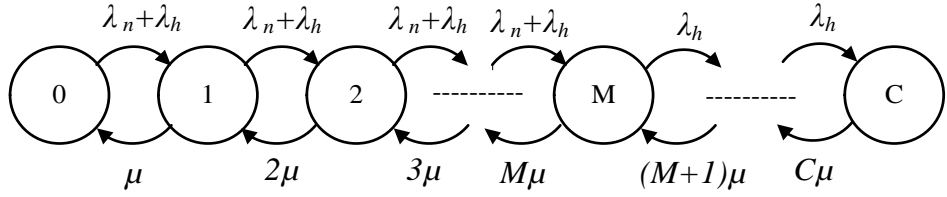

**Fig. 3.1:** State transition rate diagram of FGB scheme

The state $i$ ($i = 0, 1, \ldots C$) of a cell is defined as the number of calls in a system. Let $P(i)$ be the steady-state probability that the system is in state $i$. The probabilities $P(i)$ can be found by analysing the typical way of birth–death processes of one dimensional Markov process [9]. The relevant state transition rate diagram is shown in Fig. 3.1. This figure clarifies the birth rate of new and handover calls as well as the death rate of them by average departure rate ($\mu$). From this figure, the state balance equation can be represented as,

$$i\mu P(i) = \begin{cases} (\lambda_n + \lambda_h)P(i-1), & 0 \leq i \leq M \\ \lambda_h P(i-1), & M \leq i \leq C \end{cases} \quad (3.4)$$

In (3.4), $\lambda_n$ and $\lambda_h$ denotes the call arrival rate of new calls and handover calls, respectively. The steady-state probability $P(i)$ is found as,

$$P(i) = \begin{cases} \dfrac{(\lambda_n + \lambda_h)^i}{i!\mu^i} P(0), & 0 \leq i \leq M \\ \dfrac{(\lambda_n + \lambda_h)^M \lambda_h^{i-M}}{i!\mu^i} P(0), & M+1 \leq i \leq C \end{cases} \quad (3.5)$$

where,

$$P(0) = \left[ \sum_{i=0}^{M} \frac{(\lambda_n + \lambda_h)^i}{i!\mu!} + \sum_{i=M+1}^{C} \frac{(\lambda_n + \lambda_h)^M \lambda_h^{i-M}}{i!\mu!} \right]^{-1} \quad (3.6)$$

The blocking probability, $P_B$ for a new call is given by,

$$P_B = \sum_{i=M}^{C} P(i) \quad (3.7)$$

According to (3.5) to (3.7), the blocking probability of handover request or handover call dropping probability, $P_D$ is given by,

$$P_D = P(C) = \frac{(\lambda_n + \lambda_h)^M \lambda_h^{C-M}}{C! \mu^C} P(0) \tag{3.8}$$

Call accepting and rejecting strategies of FGB scheme are presented in Algorithm 1. Here, the accepting strategies of new calls and handover calls are classified in two different conditions. Channels from *M* to *C* are the guard band that accepts only the handover calls.

**Algorithm 1:** The call accepting and blocking strategies of FGB scheme

```
if      (NEW CALL) then
        if      (Num. of occupied channels < M)
                Accept call;
        else
                Block call;
        end if
end if

if      (HANDOVER CALL) then
        if      (Num. of occupied channels < C)
                Accept call;
        else
                Block call;
        end if
end if
```

### 3.3.2  Fractional guard channel scheme

In the FGC scheme, new calls are accepted with a certain probability that depends on the current channel occupancy which is also recognized as thinning scheme-I [19], [22]. This scheme is known as thinning scheme because the new call accepting factor becomes thinner as the channel occupancy increases. In this case, it is necessary to randomize a parameter which denotes the probability of acceptance of a new call. It should be bear in mind that both schemes accept handover calls as long as channels are available but for the new call accepting rate is limited by the acceptance factor.

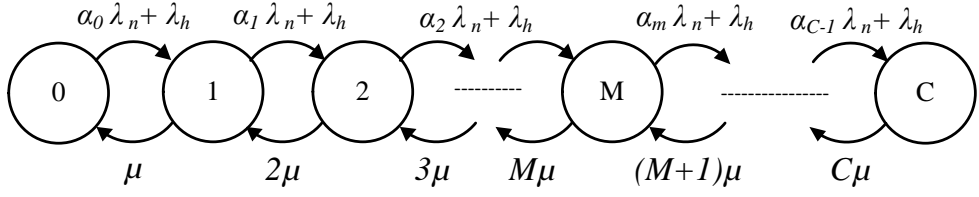

**Fig. 3.2:** State transition rate diagram of FGC scheme

The Markov process of FGC scheme is illustrated in Fig. 3.2. The state transition rate diagram of Fig. 3.2 demonstrates that form the starting state to final state the new call accepting factors are changing on the basis of channel occupancy. It starts from 1 and ends to 0. The probability of channel occupancy at any state, $i$, $P(i)$ by a new call or a handover call is given by [10], [19],

$$P(i) = \frac{\prod_{j=0}^{i}\left(\alpha_j \lambda_n + \lambda_h\right)}{i!} P(0), \qquad 0 \leq i \leq C \qquad (3.9)$$

where,

$$P(0) = \left[\sum_{i=0}^{C} P(i)\right]^{-1} \qquad (3.10)$$

The blocking probability of a new call, $P_B$ is given by,

$$P_B = \sum_{i=0}^{C} \frac{(1-\alpha_{i+1})\prod_{j=0}^{i}\left(\alpha_j \lambda_n + \lambda_h\right)}{i!} P(0), \qquad (3.11)$$

and dropping probability of a handover call, $P_D$ is given by,

$$P_D = \frac{\prod_{j=0}^{C}\left(\alpha_j \lambda_n + \lambda_h\right)}{C!} P(0) \qquad (3.12)$$

Here, $\alpha$ denotes the acceptance factor and $j$ denotes the current state. So, $\alpha_j$ denotes the acceptance factor of the current state. In this scheme, $\alpha_0 = 1$, $\alpha_C = 0$ and the others are fractional values those vary randomly between 1 and 0.

The arrived call accepting and rejecting strategies of FGC scheme are analysed in Algorithm 2. In this algorithm, a function named **rand** (0,1) is initiated which can produce any rational number randomly between 0 and 1 on the basis of occupied channels those help to accept calls at that ratio and rest of the calls are rejected.

**Algorithm 2:** The call accepting and blocking strategies of FGC scheme

**rand (0, 1)** returns a uniformly generated random number in the interval [0,1]

```
if      (NEW CALL) then
        if (rand (0, 1) ≤ αᵢ (Num. of occupied channels))
            Accept call;
        else
            Block call;
        end if
end if

if      (HANDOVER CALL) then
        if (Num. of occupied channels < C)
            Accept call;
        else
            Block call;
        end if
end if
```

### 3.3.3 Limited fractional channel scheme

The state transition rate diagram of a system with C channels is illustrated in Fig. 3.3. As the name suggests, the LFC scheme is a simplification of the more general FGC scheme about which is described earlier. In the LFC scheme, after the channel is occupied up to M, new calls are accepted with a probability $\alpha$. From states $M+1$ to $C$, handover calls are accepted and in states $0$ to $M-1$, both types of calls are accepted. Thus, the randomization in the LFC scheme is restricted to just one state as compared to the FGC scheme where randomization could potentially occur at every state. CBP and CDP of LFC scheme can be easily calculated using (3.9) - (3.11) by setting $\alpha_{m+1}=\alpha$, and the values of $\alpha_i=1$, $0 \leq i \leq M$, and $\alpha_i=0$, $M+1 < i \leq C$ [10].

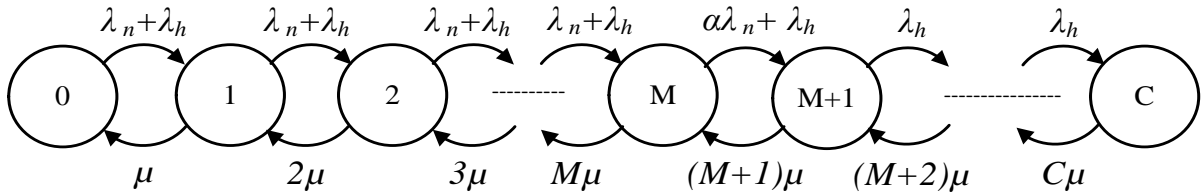

**Fig. 3.3:** State transition rate diagram of LFC scheme

The arrived call accepting and rejecting strategies of FGC scheme are presented in Algorithm 3. This algorithm is quite same as FGC scheme but in LFC scheme the acceptance factor is previously assigned. From *0* to *M* this value is 1 and from *M+1* to *C* this value is 0.

**Algorithm 3:** The call accepting and blocking strategies of LFC scheme

**rand (0, 1)** returns a uniformly generated random number in the interval [0,1]

```
if      (NEW CALL) then
        if (rand (0, 1) ≤ αi (Num. of occupied channels < M))
           Accept call;
        else if (Num. of occupied channels == M))
              AND (rand (0, 1) ≤ α)
           Accept call;
        else
           Block call;
        end if
end if
if      (HANDOVER CALL) then
        if (Num. of occupied channels < C)
           Accept call;
        else
            Block call;
        end if
end if
```

### 3.3.4 Uniform fractional channel scheme

The UFC scheme uses new call admission probability, $\alpha$ independent of channel occupancy to accept new calls. The state transition rate diagram of UFC scheme is shown in Fig. 3.4. This scheme accepts handover calls as long as channels are available. This policy can be obtained from FGC scheme by setting $\alpha_k = \alpha$, (for $k = 0, ... , C - 1$). UFC scheme reserves non-integral number of guard channels for handover calls by rejecting new calls with some probability. According to the studies given in [32]-[34] show that the UFC scheme has a lower blocking probability for new calls in low handover and new calls traffic ratio.

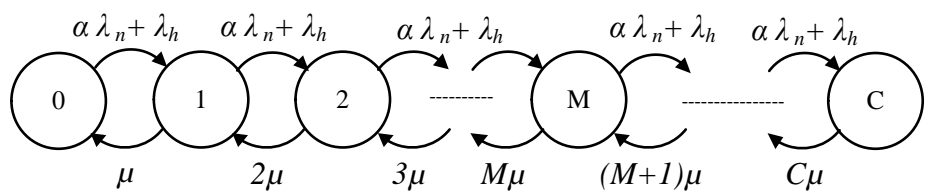

**Fig. 3.4:** State transition rate diagram of UFC scheme

The steady-state probability $P(i)$ calculating function [32] is given as,

$$P(i) = \frac{(\alpha\lambda_n + \lambda_h)^i}{i!\mu^i} P(0), \qquad 0 \leq i \leq C \qquad (3.13)$$

Here, $P(0)$ can be calculated by the equation is given as,

$$P(0) = \left[1 + \sum_{i=1}^{C} P(i)\right]^{-1} \qquad (3.14)$$

The handover CDP and new CBP of this scheme is calculated as,

$$P_D = P(C) \qquad (3.15)$$

$$P_B = (1-\alpha)\sum_{i=0}^{C-1} \frac{(\alpha\lambda_n + \lambda_h)^i}{i!\mu^i} P(0) + P(C) \qquad (3.16)$$

The arrived call accepting and rejecting strategies of UFC scheme are presented in Algorithm 4. In this algorithm a function entitled **rand** (0, 1) is also initiated which produces a uniform acceptance factor randomly regardless the channel occupancy. By this technique it is different from thinning scheme I and thinning scheme II.

**Algorithm 4:** The call accepting strategy of UFC scheme

**rand (0, 1)** returns a uniformly generated random number in the interval [0,1]

```
if      (NEW CALL) then
        if (Num. of occupied channels < C and rand (0, 1) < α) then
            Accept call;
        else
            Block call;
        end if
end if

if      (HANDOVER CALL) then
        if (Num. of occupied channels < C ) then
            Accept call;
        else
            Block call;
        end if
end if
```

## 3.4 Proposed handover priority scheme

The proposed scheme is decorated by hybridizing three set of bands. A uniform fractional band is assigned between the non-priority and priority bands of fixed guard band scheme. For this reason, this scheme is named by UFB scheme. UFB scheme can be illustrated by the given one dimensional Markov chain in Fig. 3.5.

The three bands, existed in this scheme, accept call with uniform acceptance factor by three different patterns. The first band is non-priority band where both the new calls and handover calls are accepted with same priority. In this non-priority band the acceptance factors for new calls and handover calls are 1. The second band in UFB scheme is the fractional band. In this set of channels, the new calls are accepted by a predefined acceptance factor which is less than 1 and this acceptance factor throughout the band is uniform. In fractional band the handover calls are accepted by the acceptance factors, 1. The last band is the integral priority band where the channels are reserved only for handover calls. In this case, it can be said that the new calls are accepted by this band with void acceptance factors.

States from $0$ to $M$ in Fig. 3.5, new calls and handover calls have no priority to access. When the states up to $M$, are occupied, the new calls are accepted by the states from $M+1$ to $N$, with a uniform acceptance factor, $\alpha$. This acceptance factor is independent of the channel occupancy through the band. This type of priority is termed as fractional priority. The states from $N+1$ to $C$ are reserved only for handover calls like the FGB scheme. It means these states or the corresponding set of channels accept only handover calls. So, the acceptance factors throughout the band for new calls are void. The priority of handover calls given by this band is termed as integral priority.

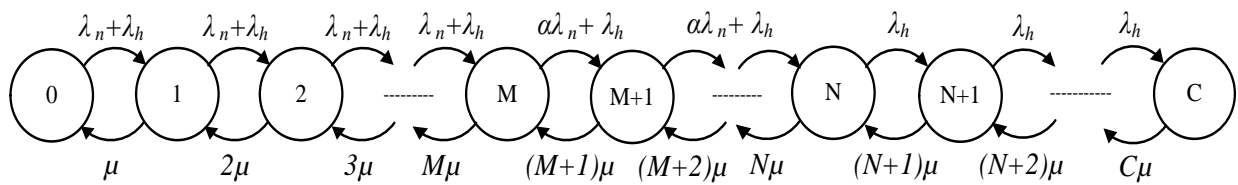

**Fig. 3.5:** State transition rate diagram of UFB scheme

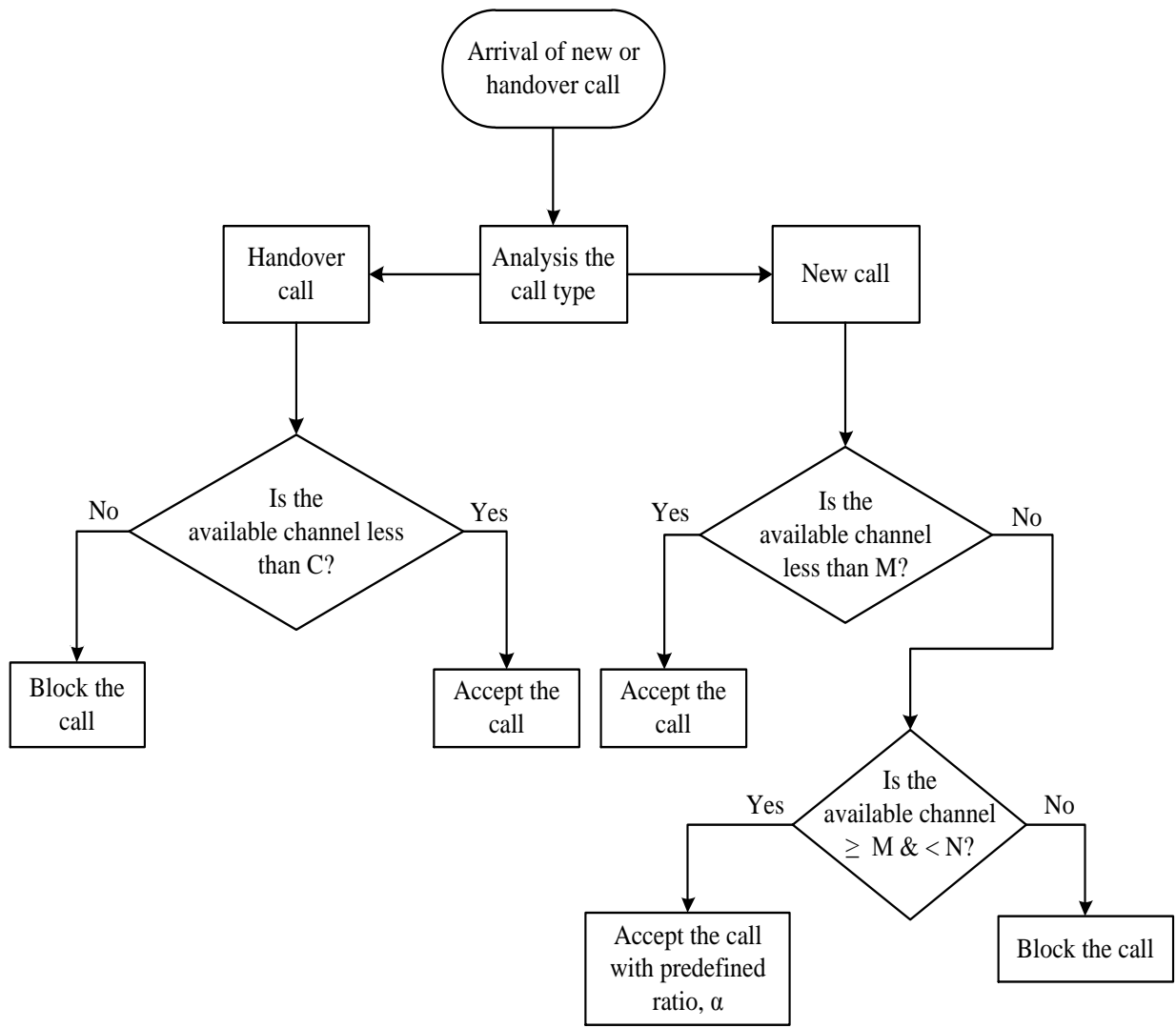

**Fig. 3.6:** Steps and conditions for call admission controlling of UFB scheme

The ideas of call accepting and blocking the new calls and handover calls of UFB scheme are presented by Fig. 3.6. At first, the network identifies the call whether it is either new call or handover call. If the call is handover call and there is a free channel in the system, the handover call is accepted by this scheme. If there is no available channel to access, handover call will be blocked. From states $0$ to $M$, all new calls are accepted but form states $M$ to $N$, new calls are accepted by predefined acceptance ratio. The priority band, from states $N$ to $C$, the channels are only reserved for handover calls which means all new calls are blocked by these states.

According to UFB scheme, priorities to the handover calls are provided by two steps. The steady state probability functions are classified into three conditions. First condition represents the non-priority characteristics, second condition presents the uniform fractional characteristics and the third condition is integral priority for the handover calls. That is why, the steady state probability, *P(i)* of UFB scheme is presented with aforesaid three conditions by,

$$P(i) = \begin{cases} \dfrac{(\lambda_n + \lambda_h)^i}{i!\mu^i} P(0), & 0 \leq i \leq M \\[2ex] \dfrac{(\lambda_n + \lambda_h)^M (\alpha\lambda_n + \lambda_h)^{i-M}}{i!\mu^i} P(0), & M \leq i \leq N \\[2ex] \dfrac{(\lambda_n + \lambda_h)^M (\alpha\lambda_n + \lambda_h)^{N-M} \lambda_h^{i-N}}{i!\mu^i} P(0), & N \leq i \leq C \end{cases} \quad (3.17)$$

According to the laws of probability we can write that,

$$\sum_{i=0}^{C} P(i) = 1 \qquad (3.18)$$

To equate (3.18) we find the value of the state occupancy probability of null state, *P(0)* which is given as,

$$P(0) + \sum_{i=1}^{M} \dfrac{(\lambda_n + \lambda_h)^i}{i!\mu^i} P(0) + \sum_{i=M+1}^{N} \dfrac{(\lambda_n + \lambda_h)^M (\alpha\lambda_n + \lambda_h)^{i-M}}{i!\mu^i} P(0) + \sum_{i=N+1}^{C} \dfrac{(\lambda_n + \lambda_h)^M (\alpha\lambda_n + \lambda_h)^{N-M} \lambda_h^{i-N}}{i!\mu^i} P(0) = 1$$

so, $$P(0) = \left[ 1 + \sum_{i=1}^{M} \dfrac{(\lambda_n + \lambda_h)^i}{i!\mu^i} + \sum_{i=M+1}^{N} \dfrac{(\lambda_n + \lambda_h)^M (\alpha\lambda_n + \lambda_h)^{i-M}}{i!\mu^i} + \sum_{i=N+1}^{C} \dfrac{(\lambda_n + \lambda_h)^M (\alpha\lambda_n + \lambda_h)^{N-M} \lambda_h^{i-N}}{i!\mu^i} \right]^{-1} \quad (3.19)$$

According to the concept of UFB scheme, the new calls are started to block after the states up to *M* are occupied and this blocking rate will be as (1 - acceptance factor). Such blocking pattern runs up to the channel occupancy of *N* state. Beyond the states *N* are occupied the new calls are blocked totally. This is why, the blocking probability of new calls can be calculated as,

$$P_B = (1-\alpha) \sum_{i=M}^{N-1} \dfrac{(\lambda_n + \lambda_h)^M (\alpha\lambda_n + \lambda_h)^{i-M}}{i!\mu^i} P(0) + \sum_{i=N}^{C} \dfrac{\lambda_h^{i-N} (\lambda_n + \lambda_h)^M (\alpha\lambda_n + \lambda_h)^{N-M}}{i!\mu^i} P(0) \quad (3.20)$$

The handover call blocking is known as handover CDP which is mentioned earlier. Since the handover calls are accepted by the total channels, a handover call will be dropped if the all channels are occupied which means all channels become busy in service. Here, if total channels are *C*, the state occupying probability of state *C* is the probability of blocking a handover call.

So, according to the third conditional function of state transition probability given in (3.17), the handover CDP can be computed as,

$$P_D = P(C)$$
$$= \frac{\lambda_h^C (\lambda_n + \lambda_h)^M (\alpha\lambda_n + \lambda_h)^{N-M}}{C! \mu^C} P(0) \quad (3.21)$$

In Algorithm 5 call accepting and rejecting strategies of UFB scheme are given. In this algorithm the function named **rand** (0, 1) produces any rational number between 0 and 1.

**Algorithm 5:** The call accepting strategy of UFB scheme
**rand (0, 1)** returns a uniformly generated random number in the interval [0,1]

```
if      (NEW CALL) then
        if (Num. of occupied channels < M)) then
           Accept call;
              else if (Num. of occupied channels => M  & < N and rand (0, 1) < α) then
                      Accept call;
              else
                      Block call;
              end if
        end if
end if

if      (HANDOVER CALL) then
        if (Num. of occupied channels < C ) then
           Accept call;
        else
           Block call;
        end if
end if
```

## 3.5   Performance analysis

In this section, the simulation results are presented for assessment of proposed scheme with the others. These results show how much deviation may be caused by using the proposed scheme, and other traditional CAC schemes. On the other hand, the various features of the proposed scheme are also described gradually.

First of all, investigation of all the CAC schemes go on considering the total channels, *C=100*, the guard band for the FGB scheme and for the limited fractional channel scheme, *M=90*, and in the proposed scheme, *N=94*. The new call arrival rate is considered from 0 to 6 calls per second in every case. Since the authors of [10], [32]-[35], consider the handover call as a fixed ratio with new call arrival rate, to compare their results with the UFB scheme in every simulation the handover call is considered as 1/6 times of the new call arrival rate. Mean call life time, *1/μ* is considered as for both new calls and handover calls as, 90 second and the mean dwell time, *1/η*=360 second.

At first, in numerical results, the new CBP's and handover CDP's of the various CAC schemes are examined. Fig. 3.7 presents such a comparison among the various popular CAC schemes as well as the proposed scheme too. This figure shows that the proposed scheme confirms the lowest new call blocking probability than the others and the highest call blocking probability is occurred in case of the FGC scheme. UFC scheme has a lower CBP than LFC, FGC, and FGB schemes at the higher new call arrival rate but very poor performance in the lower new call arrival rate.

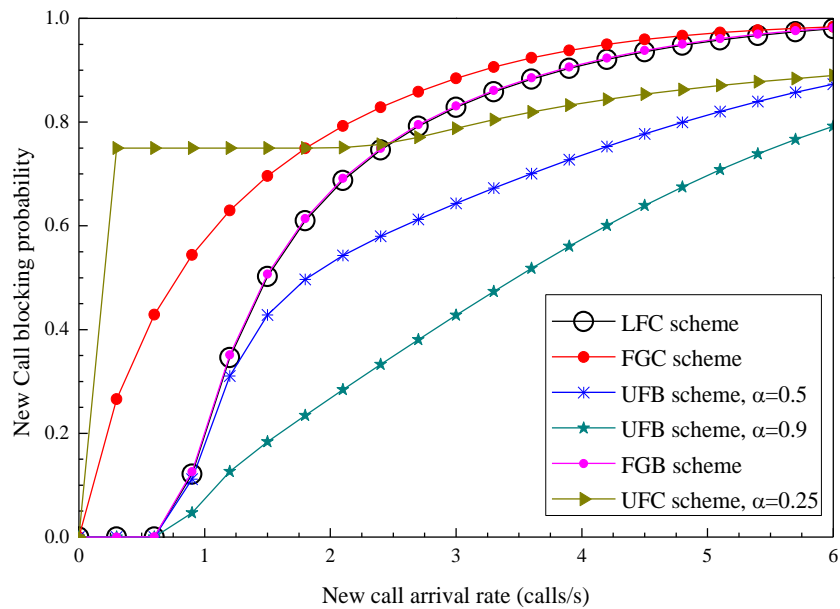

**Fig. 3.7:** Comparison of new call blocking probabilities among the conventional guard band CAC schemes with the proposed UFB scheme (new call: handover call=6:1)

The handover CDP comparisons are presented in Fig. 3.8. Here, it is observed that the handover CDP of the UFB scheme increases slightly than the LFC and FGB schemes but much more less than the UFC and FGC schemes. It is also noteworthy information in this case that the values of call acceptance factors do not impact notably, on the handover CDP of the UFB scheme while the number of fractional channels are very few (here only 4 channels are used) with respect to total channel. In this figure, two different acceptance factors 0.5 and 0.9 are considered for accepting the new calls. For these two values, the handover CDP's of UFB scheme do not vary remarkably. According to Fig. 3.8, it is clear that UFC scheme is the worst CAC scheme for the QoS aspect because its handover call dropping probabilities are much more greater than the UFB, LFC, FGC, and FGB schemes. FGC and LFC schemes show quite same results but their handover CDP's are less than the FGC and UFC schemes. Since the handover call rate is increasing with the increment of new calls (new call: handover call = 6:1), handover CDP increases drastically with the increase of handover call rate. According to this outcome, the event is truth in case of every scheme. A hypothesis on handover call rate is discussed earlier that represents that handover call rate becomes almost constant after a certain limit of new call arrival rate. Therefore, to get the proper concept about the performances of these schemes, hypothetical result on handover call rate form (3.2) is to be considered.

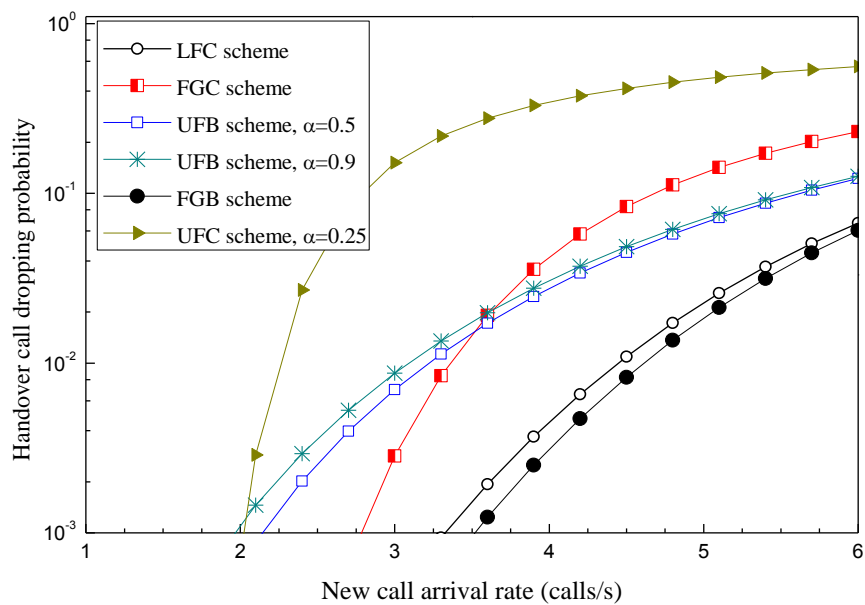

**Fig. 3.8:** Comparison of handover call dropping probabilities among the conventional guard band CAC scheme with the proposed UFB scheme (new call: handover call=6:1)

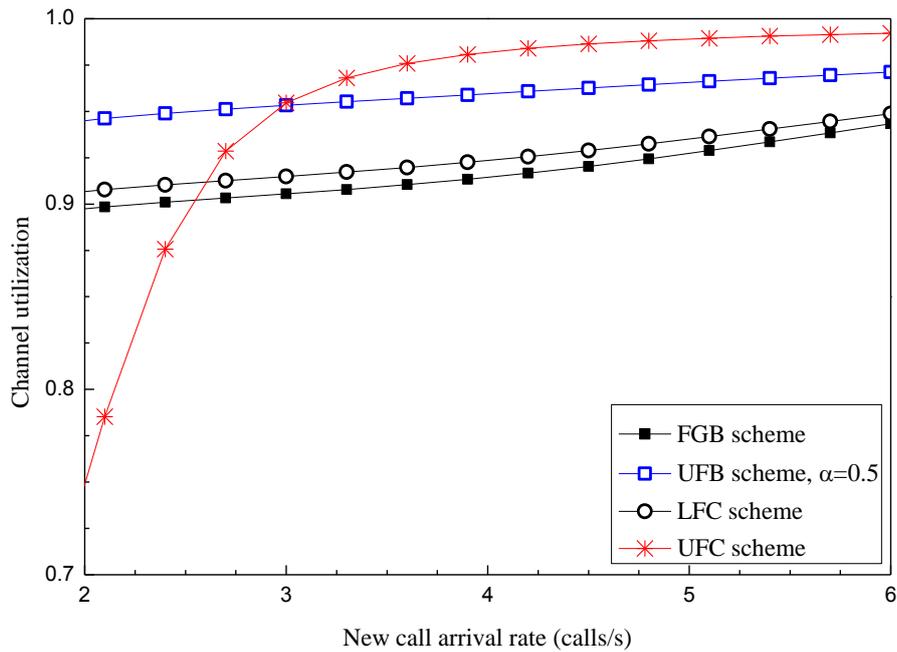

**Fig. 3.9:** Assessments of channel utilization among the various CAC schemes
(new call: handover call=6:1)

The channel utilization by the different CAC schemes are presented in Fig. 3.9. It is observed from this figure that at the higher call arrival rate UFC scheme utilizes most channels but at the lower call rate its performance in the aspect of channel utilization is the lowest one. This result states that the FGB scheme cannot assure the proper utilization of radio resources. Comparatively, beyond the UFB scheme, the LFC scheme shows the better performance in the channel utilization aspect. Eventually, in this figure it is clear that, the proposed UFB scheme utilizes more channels than the LFC scheme and obviously this performance maintains its consistency through the lower to higher new call arrival rate. Thus, UFB scheme illuminates the limitation of UFC scheme because UFC scheme is not consistent in its channel utilization performance from lower to higher call arrival rates. From the Fig. 3.9, it is also noticeable that, though the handover call dropping probabilities of UFB scheme are slightly higher than FGB and LFC schemes, it utilizes more channels than FGB and LFC scheme.

Overall CBP is one of the performance analytic measurements. With the augmentation of overall CBP, the system cost increases. To reduce the system cost, overall CBP is a concerning issue. A comparison of overall call blocking probabilities of UFC, UFB, FGC, FGB, and LFC schemes are represented by Fig. 3.10. In lower traffic, the overall call blocking probabilities of UFC scheme are greater than the other schemes but at higher traffic UFC performs better in aspect of overall call blocking probabilities than FGB, LFC, and FGC schemes. Since the CBP of the proposed UFB scheme is lowest than the other schemes as well as its handover CDP increases very slightly, the overall CBP of the UFB scheme shows the lowest overall CBP. Here, it is also observed that in the concern of overall CBP, UFC and FGC schemes show very poor performances. From this figure, it is clear that UFB scheme is the best scheme in the aspect of overall call blocking probability with respect to UFC, FGB, FGC, and LFC schemes.

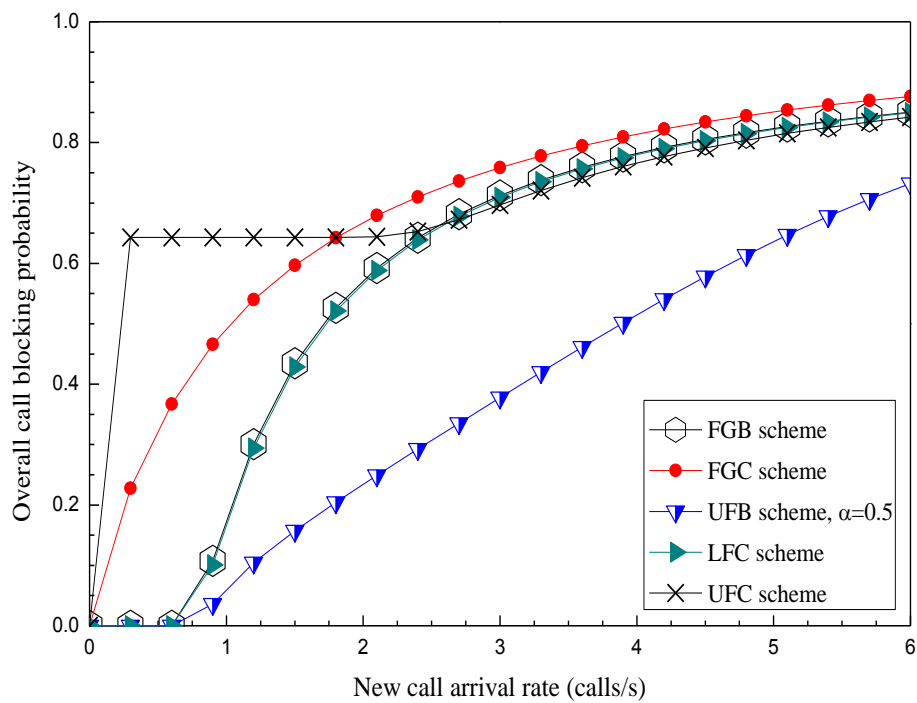

**Fig. 3.10:** Valuation of overall call blocking probabilities of the different CAC schemes
(new call: handover call=6:1)

Now, some special factors of the proposed scheme will be discussed. Most authors [6]-[11] have been considered the handover call rate as a constant function of new call arrival rate. According to the proposition in [3], [11] the handover call arriving rate follows (3.2). This relation depicts that handover call rate becomes almost constant after a certain limit of new call arrival rate. So, the numerical performance of UFB, UFC, FGB, FGC, and LFC scheme are analyzed considering (3.2). For this reason, estimations of the rate of handover call by the different CAC schemes are illustrated in Fig. 3.11. From this figure, it is observed that the handover call rate of UFB, LFC, and FGB schemes become almost constant after the new call arrival rate increases at 1.5 calls/s. In this figure it is also observable that the handover call rate of UFB scheme is slightly less than the FGB and LFC scheme. This relation assigns that considering the handover call rate as a constant function of new call arrival rate may underestimate or overestimate the new CBP and handover CDP of any CAC scheme.

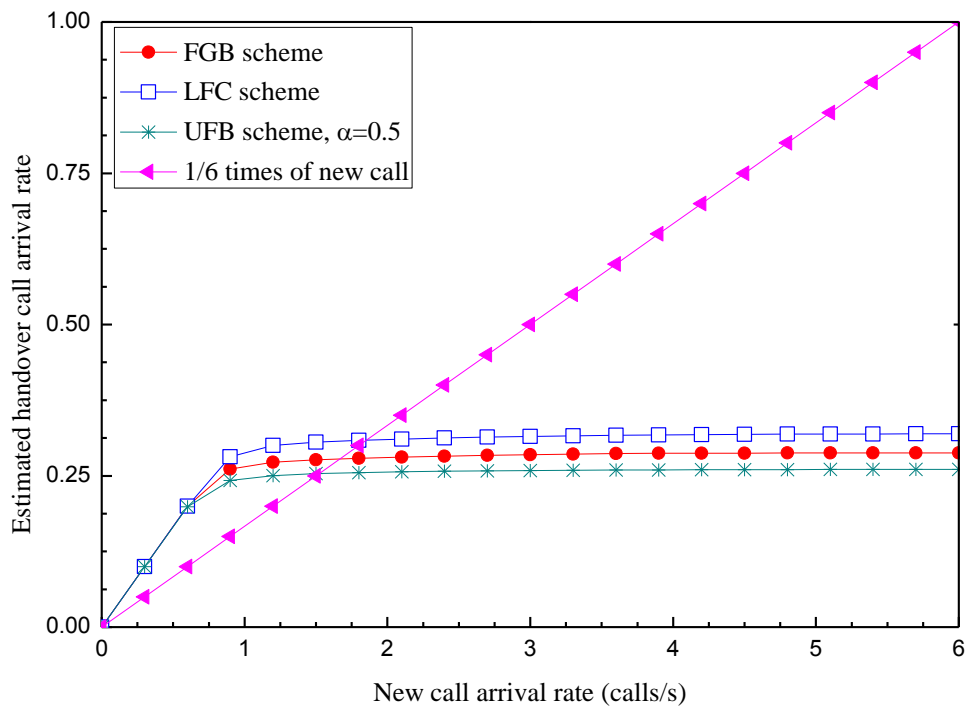

**Fig. 3.11:** Estimation of handover call arrival rate by statistical analysis

Wrong estimation of the handover call rate has negative impacts on new CBP and handover CDP because the total call arrival rate (handover call rate + new call rate) is changed. To observe these impacts on new CBP and handover CDP of different CAC schemes, Fig. 3.12 and Fig. 3.13 are presented. In Fig. 3.12 the handover CDP of fixed and proposed conditions for different CAC schemes are presented. In Fig. 3.12 it is observed that the handover CDP is overestimated in fixed ratio with new call arrival rate. By the application of proposed rate of handover call rate, FGB scheme and UFB scheme shows almost same handover call dropping probability. UFB scheme also shows lower handover CDP than that of FGC scheme. This result ensures that UFB scheme is no more threat for QoS which was a concerning issue in the previous result of fixed handover call ratio with new call arrival rate. In Fig. 3.13 the new CBP of different schemes after applying the proposed scheme is presented where it is observed that the new CPB of the UFB scheme is lower than the FGC, FGB, and LFC schemes. So, from Fig. 3.12 and Fig. 3.13 it can be concluded that UFB scheme is the most efficient handover priority scheme because it offers the lowest new CBP providing the proper QoS.

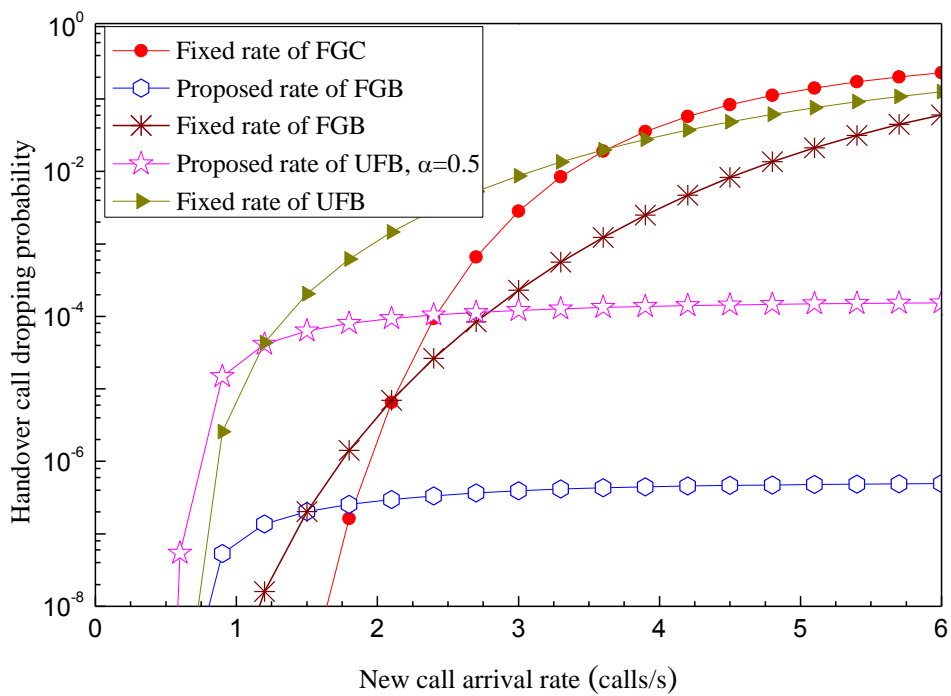

**Fig. 3.12:** Comparison of handover CDP between the proposed and fixed call handover rate

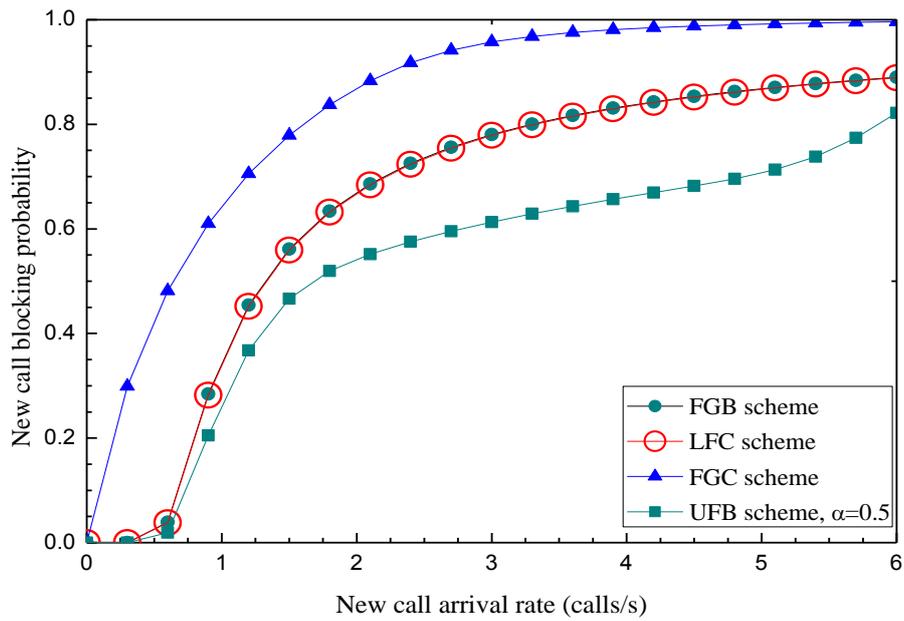

**Fig. 3.13:** New CBP of FGB, LFC, FGC, and UFB schemes at the statistical rate of handover call

The proposed UFB scheme is consist of two bands where new call blocking is occurred. The first band is recognized as fractional blocking band, and the second is guard band. The blocking patterns of these two bands are not similar. With the augmentation of new call arrival rate, the blocking pattern of the guard band increases, but for the fractional blocking band it increases to a certain limit and then decreases. Such a pattern is shown in Fig. 3.14. In this figure, the overall new CBP of UFB scheme is also obtainable.

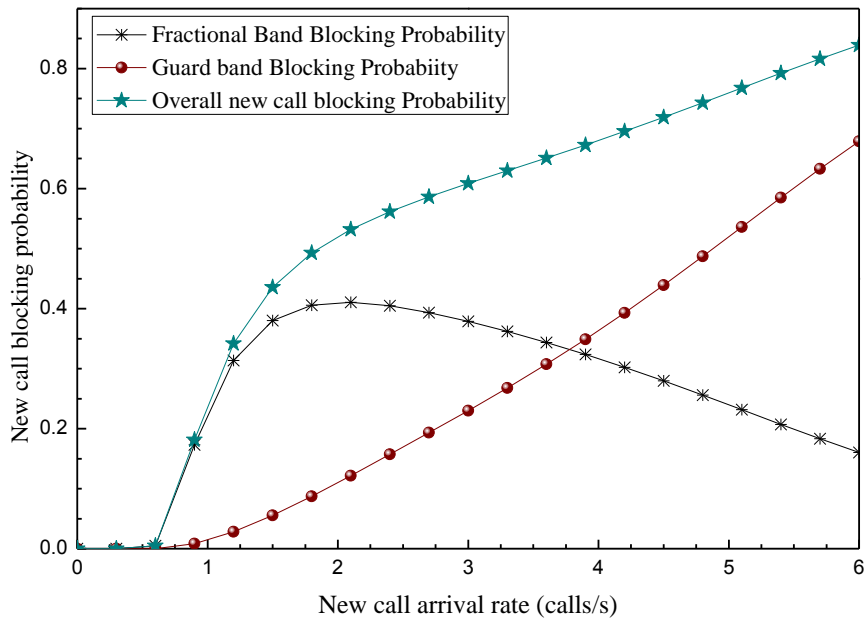

**Fig. 3.14:** Blocking pattern of new calls by the fractional blocking band and guard band

The new CBP of the proposed scheme depends on the acceptance factors. The acceptance factor and the new CBP maintain a reverse relation. Such a relation is shown in Fig. 3.15. In this figure, it is observed that as the acceptance factor increases, the new CBP decreases. Here, there a problem arises for the handover CDP. As the acceptance probability of new call increases the CDP may increase, which is a threat for maintaining the QoS of the network. In this case, the fact of hope that, if the handover call rate is very less than the new call arrival rate, the handover CDP remains almost constant. Fig. 3.16 shows that the analytical handover CDP remains almost constant with considering the statistical probability of handover call rate. In practice, the handover call rate is really much more less than the new call arrival rate. This result removes the anxiety of the aforementioned threat for maintaining the QoS of the wireless cellular networks.

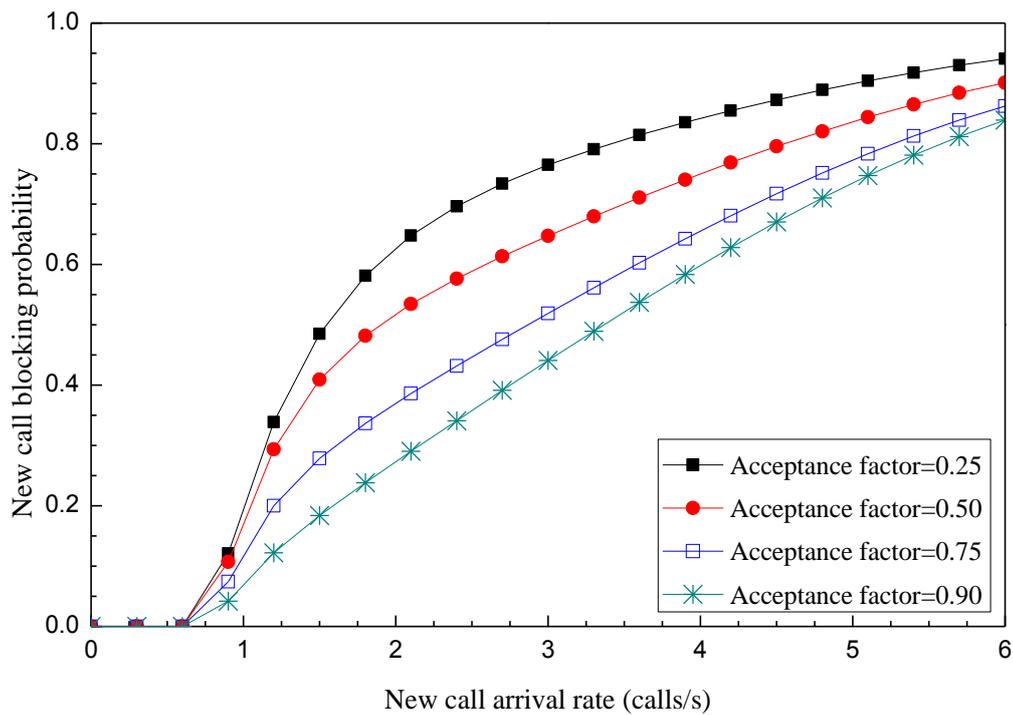

**Fig. 3.15:** New call blocking probabilities for different acceptance factors

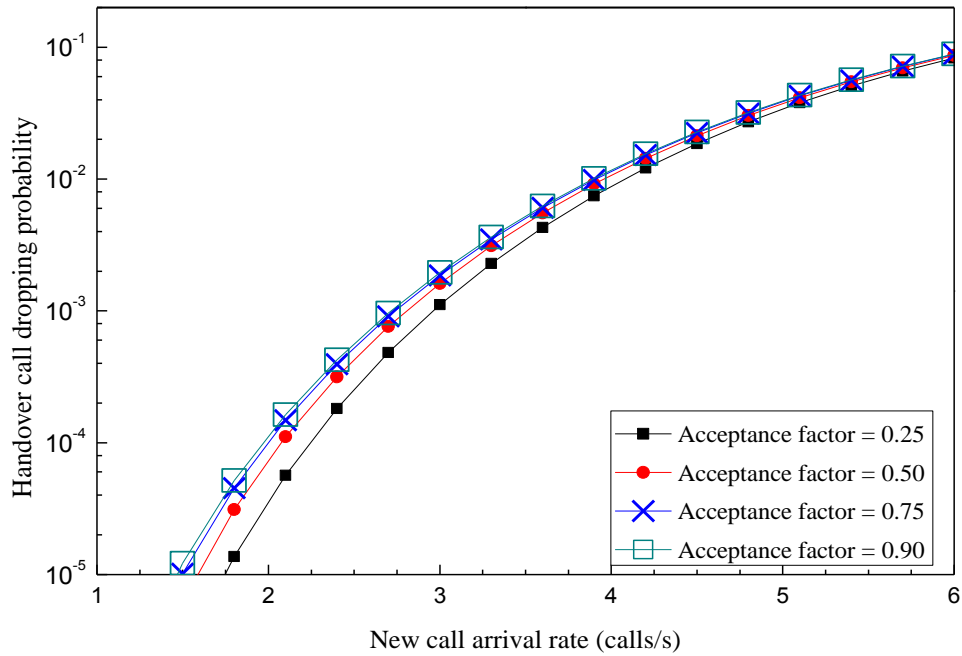

**Fig. 3.16:** Handover call dropping probabilities for different acceptance factors

## 3.6   Summary

In this section, the total research work about the handover priority scheme and the outcomes are concluded briefly as well as suggestions for implanting the proposed scheme are also discussed. The different CAC schemes based on fractional guard channel and their performance measuring factors are derived numerically. In addition with them, a CAC scheme i.e. uniform fractional band scheme is proposed. The mathematical equations of the various performance measuring factors are derived. The numerical performances of the existing CAC schemes as well as the proposed scheme are presented graphically to compare their performances. These figures, separately, describe the performance of the proposed scheme where and why special and better than the other existing schemes.

By the proposed scheme, it is shown that the performances of the scheme are better than UFC and LFC scheme. For higher traffic rate, the proposed scheme maintains its consistent characteristics too. Besides, the new CBP and channel utilization of the proposed scheme is also better than the LFC and GFB scheme. In this circumstance, there is a concern about handover CDP, because the handover CDP of the proposed scheme may increase slightly than LFC and FGB scheme based on the value of acceptance factor. This is why the statistical analysis for handover call rate is described with respect to the new call arrival rate considering the blocking and dropping probability of the system instead of the fixed rate of new call arrival rate. This analytical proposition clears that the system faces the call handover rate is different for different CAC schemes. So, in the case of underestimation or overestimation of the call handover rate does not show the same handover CDP. This consideration shows that the proposed scheme maintain the CDP in the satisfactory level which ensures the QoS of the network. The proposed scheme also ensures the lowest overall CBP that reduces the system cost.

From the simulation results discussed so far, it can be claimed that the proposed scheme is the most efficient CAC scheme among the existing CAC schemes in terms of CBP, CDP, channel utilization, and overall CBP as well as the system cost. As a result, this scheme mostly optimizes the QoS. So, the proposed UFB scheme can be effectively used in very higher traffic oriented wireless networks.

# CHAPTER 4

UBT scheme for multiservice networks

Chapter Outlines

- ❖ Introduction
- ❖ FGB scheme for multi-class
- ❖ UBT scheme
- ❖ Performance analysis
- ❖ Summary

## 4.1 Introduction

Communication revolution becomes possible by the incredible advancements of cellular communication. Enormous number of users belongs to the modern wireless cellular networks. Wireless system faces a huge traffic by reason of providing integrated services, such as the voice, data, and different types of multimedia. The demand for multimedia services over the air is increasing day by day which leads to design consideration of wireless internet of wireless cellular network. These different types or classes of traffic are not equally important in the aspect of service. Traffic like security, healthcare, banking, handover calls, etc. are more important where non-real time calls like data, voice messaging etc. are comparatively less important. So, these different types of traffic are to divide into several classes.

During the resource management, the important classes of traffic are prearranged higher priority and comparatively less important calls are considered as lower priority. Blocking a lower priority call is preferred over blocking a higher priority call for maintaining the QoS. In order to maintain such miscellaneous service requests of multiple traffic classes using limited resources, efficient resource management and QoS provisioning is very important issue [33], [36], [39]. CAC is such a provisioning technique to maintain the QoS among the several traffic classes by limiting the number of call connections into networks that reduces the blocking probability of higher priority traffic classes [5] and also increases the channel utilization.

A number of CAC schemes have been proposed in [1], [2], [5], [19], [20], [34]-[36] which endow with different level of priorities among different traffic classes for maintaining the QoS. The schemes [5], [35], [40], [41] are proposed based on the notion of FGB or QoS adaptability. A number of channels are reserved for exclusive use by a particular class of traffic in a FGB scheme. This is very easy mechanism to reduce the call blocking probability of higher class of traffic but this type of CAC schemes decrease the channel utilization. Adaptive bandwidth based schemes [3], [36], [42, 43] reduce resource allocation for ongoing calls to accept more calls of higher priority. Class-based QoS provisioning is done by flexible resource allocation in [35] where optimum channel utilization is not considered. In [33] a mechanism is described by which blocking probability of only one class can be reduced providing constraints to the other traffic classes but in this work there is no clear conception about the channel utilization. As a result, both the QoS and optimum utilization of resources are necessary to consider for designing the CAC scheme that reduces the blocking probability of lower priority traffic maintaining the blocking probability of higher priority traffic at a reliable level.

In this chapter, a new idea about CAC scheme has been proposed based on FGB scheme and uniform thinning scheme [32]. In thinning schemes, any traffic class can be accepted by a predefined probability on the basis of the present channel occupancy [10] or the amount of call arrival rate [19], [22]. In the aspect of multiclass traffic the call admission techniques become very complex if the total channels are accessed as thinning procedure. In [19], the author first approaches about thinning scheme and generalised this scheme for multiclass by mathematical modelling but there is no numerical analysis of thinning scheme. In FGB scheme, there are multiple thresholds for multiple traffic classes. The highest priority class is accepted by the total channels. Some channels are reserved for all classes according to their priorities. The reserved channels are called guard band. The arrival request of adjacent priority class is terminated by the guard band [8]. For this channel reservation in FGB scheme, it reduces the channel utilization.

In this scheme, an idea of using the uniform thinning technique (UTT) inside every band is proposed. In FGC scheme or thinning scheme, the fractionizing of the guard obeys the decreasing manner (form 1 to 0) throughout the guard band. UTT fractionises a band of channels with a constant fraction rate. Due to accept some more calls, the blocking probability of lower priority traffic decreases, and increases the channel utilization, but it may increase the blocking probability of higher classes traffic due to accept the calls at inexact acceptance factors. Inexact acceptance factors can reduce the QoS of the system. This is why, this CAC scheme has been designed considering the blocking probability of the higher priority traffic class unaffected by fractionizing the bands properly.

In UBT scheme prohibited traffic class of FGB gets access by the corresponding band with a certain acceptance factor independent of channel occupancy. By this differentiating idea, it has been shown that the CBP of lower priority traffic class can be reduced significantly by maintaining the optimum QoS. The novelty of this CAC scheme is to increase the channel utilization and decrease the overall CBP than FGB. Furthermore, a comparison is presented with this proposed scheme, FGB scheme, and non-priority scheme. It is also described the impact of different call acceptance factor on CBP of different traffic and the benefit of UTT putting into operation.

## 4.2  FGB scheme for multi-class

The traffic faced by the cellular network is practically different types such as voice call, data, multimedia, etc. Among all traffic, classes are to define depending upon their priorities. The lower priority calls are generally blocked more than the higher priority calls. This kind of QoS is easily achieved by the fixed guard band CAC scheme. In this scheme, the lower priority classes are permitted to access fewer channels than the higher one and those channels are called guard band for that class which is graphically represented by the Fig. 3.1.

Suppose that, the arrival rate of traffic classes are defined as $\lambda_1, \lambda_2, \lambda_3$, and so on and here the 1, 2, 3 are the notations for the class numbers. The channel holding time is generally considered as average and let it be $1/\mu$. Then, the traffic load is defined as the ratio of traffic arrival rate and the mean channel holding time [44] such as traffic load of class 1 is, $\rho_1 = \lambda_1/\mu$, for class 2 is, $\rho_2 = \lambda_2/\mu$, and so on. Some authors have suggested to use the traffic load instead of call arrival rate as in [19], [22], [32], [33], & [45]. Alike the handover priority scheme in this chapter traffic load is used instead of call arrival rate. Inasmuch the channel holding time is exponentially distributed, the queuing analysis can be done by considering the system as *M/M/C/C* of one dimensional Markov process. Let, $P(i)$ be the steady-state probability that the system is in state, $i$. The state balance equations for the different guard bands [1] are given by,

$$P_1(i) = \frac{1}{i!}\left(\rho_1 + \rho_2 + \ldots + \rho_m\right)^i P(0), \quad 0 \leq i \leq C_m \tag{4.1}$$

$$P_2(i) = \frac{1}{i!}\left(\rho_1 + \rho_2 + \ldots + \rho_m\right)^{C_m} \left(\rho_1 + \rho_2 + \ldots + \rho_{m-1}\right)^i P(0), \quad C_m \leq i \leq C_{m-1} \tag{4.2}$$

$$P_3(i) = \frac{1}{i!}\left(\rho_1 + \rho_2 + \ldots + \rho_m\right)^{C_m} \left(\rho_1 + \rho_2 + \ldots + \rho_{m-1}\right)^{C_{m-1}} \left(\rho_1 + \rho_2 + \ldots + \rho_{m-2}\right)^i P(0), \quad C_{m-1} \leq i \leq C_{m-2} \tag{4.3}$$

$$P_j(i) = \frac{P(0)}{i!} \prod_{k=1}^{j-1}\left(\sum_{l=1}^{m-k+1} \rho_l\right)^{C_{m-k+1}} \left(\sum_{l=1}^{m-j+1} \rho_l\right)^i, \quad C_{m-j+2} \leq i \leq C_{m-j+1} \ \& \ 1 \leq j \leq m \tag{4.4}$$

$$\text{where,} \quad P(0) = \frac{1}{1 + \sum_{i=1}^{C_1} P(i)} \tag{4.5}$$

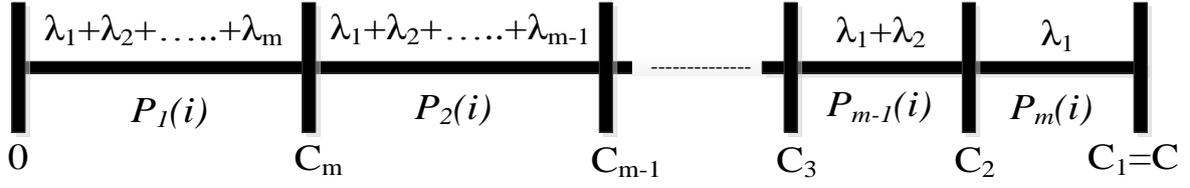

**Fig. 4.1:** Channel allocation for different classes of traffic in FGB scheme

From (4.1)-(4.4), $C$ denotes the total number of channels and $m$ denotes the number of classes. The suffix of $P$ such as 1, 2, 3, etc. denotes the corresponding guard band, respectively. We get the blocking probability of traffic class 1 by analysing (4.4) and that is defined as,

$$PB_1 = P(C_1) = \frac{P(0)}{C_1!} \rho_1^{C_1} \prod_{k=2}^{m}\left(\sum_{i=1}^{k}\rho_i\right)^{C_{k-1}-C_k} \qquad (4.6)$$

As the definition of $j$ (here, $2 \leq j \leq m$) the blocking probabilities of other traffic classes can be deduced by,

$$PB_j = \frac{P(0)}{(C_{j-1}-1)!}\left(\prod_{k=2}^{m}\left(\sum_{l=1}^{k}\rho_l\right)^{C_{k-1}-C_k}\right) \times \sum_{i=C_j}^{C_{j-1}-1}\left(\sum_{l=1}^{C_{j-1}}\rho_l\right)^{i-C_j} + \sum_{i=1}^{j-1} PB_i \qquad (4.7)$$

## 4.3 UBT scheme

In this proposed scheme, the basic idea of fixed guard band CAC scheme and uniform thinning technique inside the every band has been hybridized. UTT is an admission process that can limit the acceptance of the arrived traffic at a predefined factor or acceptance factor. The forbidden band between two consecutive classes is chosen for uniform fractionizing admission of calls of that forbidden traffic class throughout the channels of that guard band. For this reason, the proposed scheme is termed by UBT scheme. Suppose that, a band $C_{j+1}$-$C_j$ permits to accept traffic $\rho_1+\rho_2+\rho_3+....+\rho_j$. According to the FGB scheme, the next guard band, $C_j$-$C_{j-1}$ permits to accept traffic $\rho_1+\rho_2+\rho_3+....+\rho_{j-1}$. On the other hand, according to the proposed scheme in this thesis paper, that band permits to accept the traffic $\rho_1+\rho_2+\rho_3+......+\rho_{j-1}+\alpha_u\rho_j$. Here, $\alpha_u$ is the predefined acceptance factor of the corresponding guard band. In this case, by the rate of $\alpha_u$, the corresponding band is uniformly fractionized.

Due to apply this band thinning technique, the traffic load increases which is the major concern for the higher and lower priority traffic. In this case, the blocking probabilities of higher priority classes will be increased but the blocking probabilities of lower priority classes will be decreased. In fact, this is the concerning issue for QoS. Another important feature of this scheme is to choose the optimal set of acceptance factors for the corresponding bands that decrease the blocking probability, increases channel utilization maintaining the QoS. For this reason, an algorithm is necessary to apply for finding the best sets of acceptance factors for the different bands. The goal of choosing the set of acceptance factors is to reduce blocking the lower priority calls by maintaining blocking probabilities of the higher priority calls in suitable range. Some sets of acceptance factors may decrease the blocking probabilities of lower classes of traffic but increase the blocking probabilities of other higher traffic classes. According to this scheme, this kind of sets will be avoided.

FGB scheme is the easy way to gain the QoS but it shows poor performance in proper channel utilization which is overcome by our proposed UBT. The proposed scheme is designed in such a way that it is possible to reassign to FGB scheme by considering the value of acceptance factor as zero.

The total channel allocation procedure is explained by Markov chain of *K/K/C/C* queue in Fig. 4.2. Here, *K* denotes that the call arrival process is *Poison Distributed*.

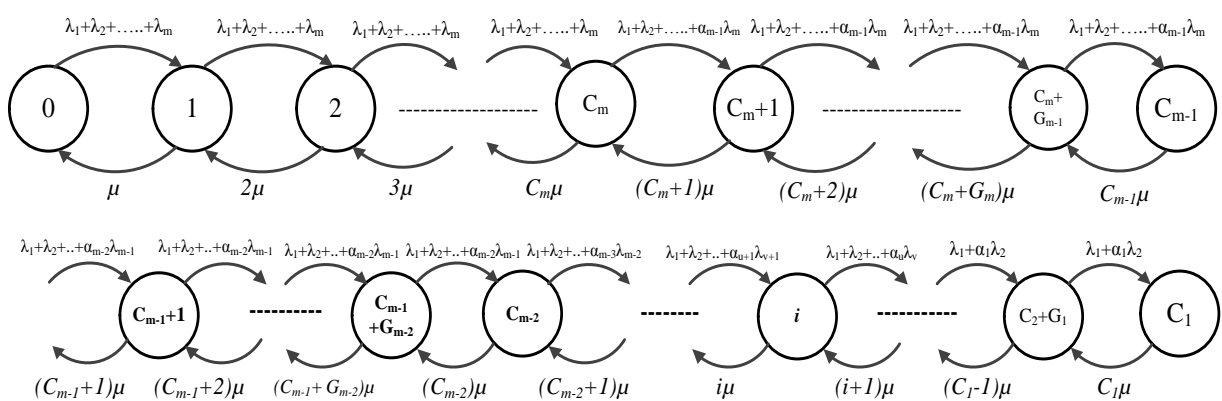

**Fig. 4.2:** State transition rate diagram of the proposed UBT scheme

The call admission process is described by the block diagram in Fig. 4.3. In this figure, $G$ is the corresponding guard length. Let, $P_j(i)$ be the steady-state probability that the system is in state, $i$ where suffix, $j$ denotes the corresponding guard band. The steady-state probability of first band i.e., $0 \leq i < C_m$ is denoted by $P_1(i)$ and is given by,

$$P_1(i) = \frac{P(0)}{i!}\left(\sum_{l=1}^{m}\rho_l\right)^i \tag{4.8}$$

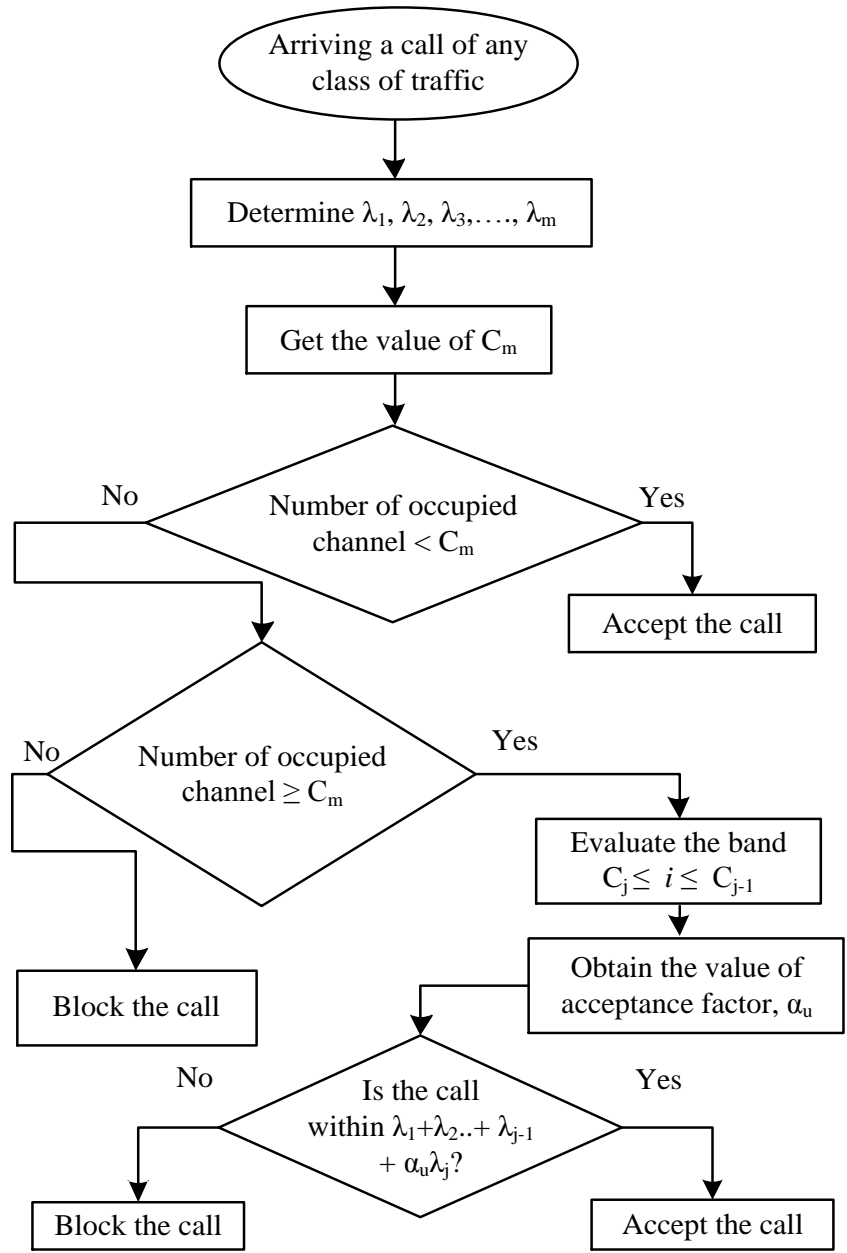

**Fig. 4.3**: Functions of estimating the bands, acceptance factors, and blocking patterns of different classes according to UBT scheme

It is clear that, uniformly band fractionizing is not done from *0* to $C_m$. That means this band is accessible for all traffic classes with same priority or the notion of same acceptance factor according to the literature of the proposed scheme. For $2 \leq j \leq m$, state occupancy probability can be evaluated by,

$$P_j(i) = \frac{P(0)}{i!} \left( \sum_{\substack{l=1, p=m-j+1 \\ q=m-j+2}}^{m-j+1} (\rho_l + \alpha_p \rho_q) \right)^i \times \prod_{k=1}^{j-1} \left( \sum_{\substack{l=1, u=m-k+1, \\ v=m-k+2}}^{m-k+1} (\rho_l + \alpha_u \rho_v) \right)^{C_{m-k+1} - C_{m-k+2}}, \quad C_{m-j+2} \leq i \leq C_{m-j+1} \quad (4.9)$$

The blocking probability of the traffic class 1 is $PB_1$ can be evaluated by,

$$PB_1 = \frac{P(0)}{C_1!} (\sum_{l=1}^{m} \lambda_m)^{C_m} \prod_{k=1}^{m-1} \left( \sum_{\substack{l=1, v=k+1, \\ u=k}}^{k} (\rho_l + \alpha_u \rho_v) \right)^{C_k - C_{k+1}} \quad (4.10)$$

In (4.9) and (4.10), it is considered that $C_{m+1} = 0$, $\lambda_{m+1} = 0$, and $\alpha_0 = 0$.

As the definition of *j*, $2 \leq j \leq m$ the blocking probabilities of other classes can be deduced by,

$$PB_j = \begin{cases} (1-\alpha_{m-j+1})\dfrac{P(0)}{i!} \sum_{i=C_{j+1}+1}^{C_j+1} \left[ \sum_{\substack{l=1, v=j \\ u=j-1}}^{j-1} (\rho_l + \alpha_u \rho_v) \right]^{i - C_{j+1}} \times \\ \prod_{k=2}^{m-1} \left( \sum_{\substack{l=1, v=k+1 \\ u=k}}^{j} (\rho_l + \alpha_u \rho_v) \right)^{C_{k+2} - C_{k+1}} + P_{pfact} + PB_1 \end{cases} \quad (4.11)$$

where,

$$P_{pfact} = \sum_{t=1}^{j-1} \left( \sum_{C_{t+1}}^{C_t - 1} \left[ \frac{P(0)}{i!} \prod_{k=1}^{j-1} \left( \sum_{\substack{l=1, u=m-k, \\ v=m-k+1}}^{m-k+1} (\rho_l + \alpha_u \rho_v) \right)^{C_{m-k+2}} \left( \sum_{\substack{l=1, p=m-j+1 \\ q=m-j+2}}^{m-j+1} (\rho_l + \alpha_p \rho_q) \right)^i \right] \right) \quad (4.12)$$

From (4.8)-(4.12) it is clear that, if the value of *α* becomes zero the system will be same as the system of FGB scheme.

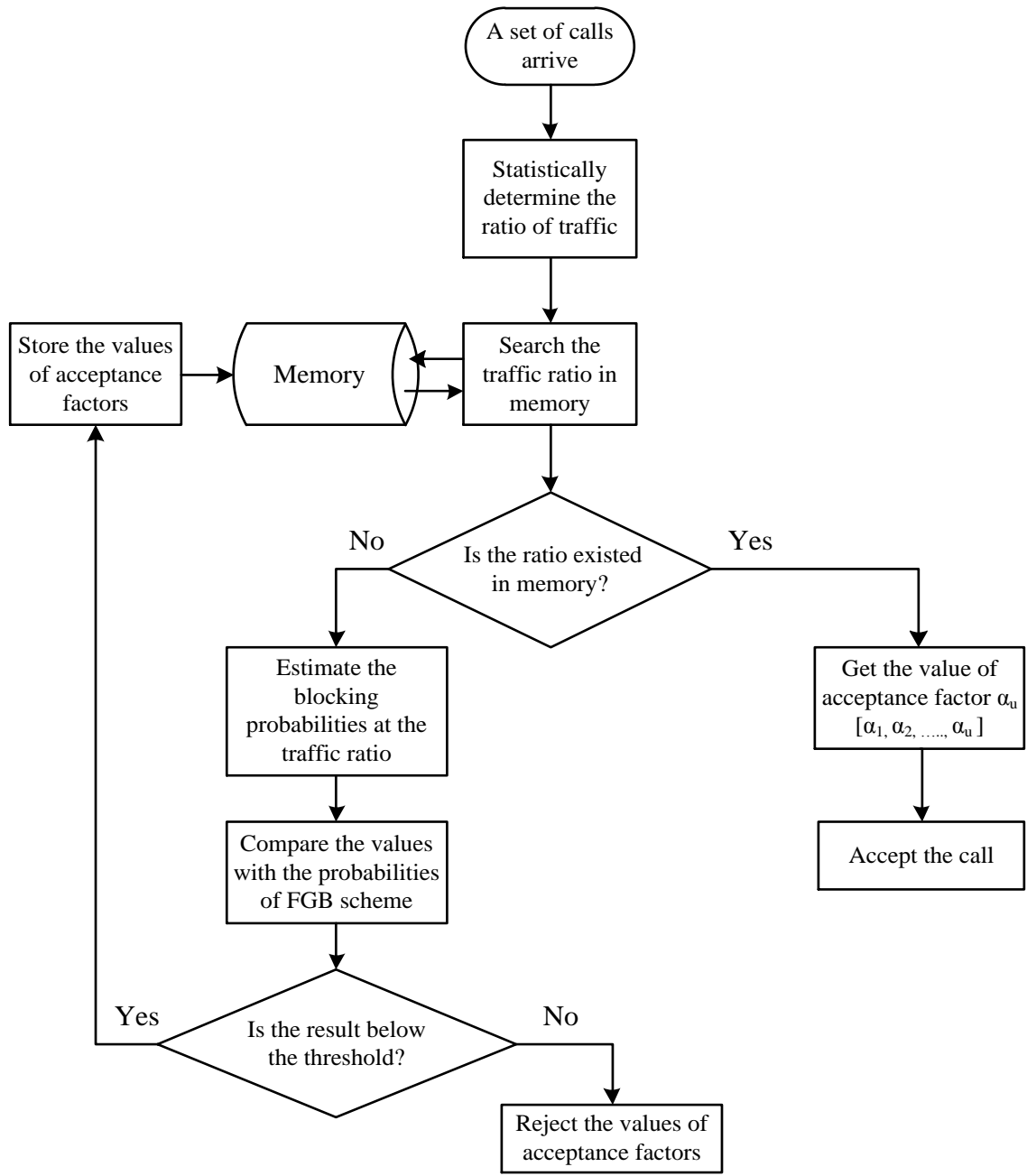

**Fig. 4.4:** The proper set of acceptance factors finding procedure for different call arrival rate to optimize the QoS

Fig. 4.4 describes a sub process of getting the set of acceptance factors. The proper sets of acceptance factors are very necessary to find the optimum service quality. In Fig. 4.4, the block diagram shows a process by which the optimum set of acceptance factors finding process is described. In this block diagram, there is a memory that store maximum probable optimum set of acceptance factors on the basis of traffic ratio. This block diagram is performed as self-data gathering process. Self-data gathering process works as,

- ➢ At first, it searches in memory to find the required data for availability.
- ➢ If there is no available required data it iteratively calculates the probable results.
- ➢ Then compare the results with QoS guaranteed threshold.
- ➢ Then select the best set of acceptance factor.
- ➢ Then store it in memory for the further usage.

When statistically the memory will be rich and learned with probable all sets of required data then the total network will get the availability of the sets of acceptance factors as required.

It is mentioned that this proposed scheme improves the channel utilization. Channel utilization depends on the blocking probabilities of different traffic classes. This measurement is necessary to realize the quantity of utilization of radio resources. According to [39], [46] the channel utilization is calculated by,

$$U_{ch} = \frac{\sum_{i=1}^{m} \lambda_i (1 - PB_i)}{\mu C} \tag{4.13}$$

In (4.13), $U_{ch}$ denotes the channel utilization and $m$ is the number of classes. In this equation, it should be taken in mind that $C=C_1$ is the total number of channels.

Else, another performance analytic parameter is overall call blocking probability. Due to provide the higher priority to some traffic classes, blocking probabilities of lower priority traffic classes of higher rate are increased. This event can increase the overall blocking probability with respect to traffic arrival rate. Overcall call blocking probability can be figured out by,

$$PB_{overall} = 1 - \frac{\sum_{i=1}^{m} \lambda_i (1 - PB_i)}{\sum_{i=1}^{m} \lambda_i} \tag{4.14}$$

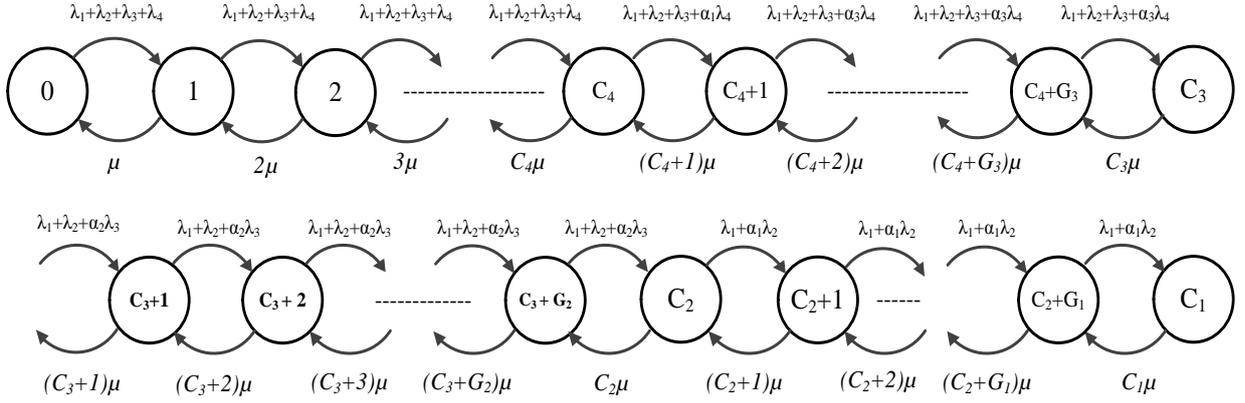

**Fig. 4.5:** State transition rate diagram of proposed UBT scheme for four traffic classes

## 4.4   Performance analysis

The proposed uniform band thinning scheme and the conventional FGB scheme as well as non-priority scheme are analysed considering average channel holding time, $1/\mu=120$ second. Total number of channels, $C$ in both cases is taken as 120. Else, the total calls are classified into four traffic classes (the value of m is 4). Since we consider that the available traffic classes are four i.e., $m=4$, the state transition rate diagram for this condition will be as Fig. 4.5. Consequently, the blocking probabilities of four traffic classes of the UBT scheme are given as,

$$PB_1 = \frac{P(0)}{C_1!}(\rho_1+\rho_2+\rho_3+\rho_4)^{C_4}(\rho_1+\rho_2+\rho_3+\alpha_3\rho_4)^{C_3-C_4}$$
$$\times (\rho_1+\rho_2+\alpha_2\rho_3)^{C_2-C_3}(\rho_1+\alpha_1\rho_2)^{C_2-C_1} \qquad (4.15)$$

$$PB_2 = \frac{P(0)}{C_2!}(\rho_1+\rho_2+\rho_3+\rho_4)^{C_4}(\rho_1+\rho_2+\rho_3+\alpha_3\rho_4)^{C_3-C_4}$$
$$\times (\rho_1+\rho_2+\alpha_2\rho_3)^{C_2-C_3} \times (1-\alpha_1)\sum_{i=C_2}^{C_2+G_1}(\rho_1+\alpha_1\rho_2)^{i-C_1} + PB_1 \qquad (4.16)$$

$$PB_3 = \frac{P(0)}{C_3!}(\rho_1+\rho_2+\rho_3+\rho_4)^{C_4}(\rho_1+\rho_2+\rho_3+\alpha_3\rho_4)^{C_3-C_4}$$
$$\times (1-\alpha_2)\sum_{i=C_3}^{C_3+G_2}(\rho_1+\rho_2+\alpha_2\rho_3)^{i-C_3} + PB_2 \qquad (4.17)$$

$$PB_4 = \frac{P(0)}{C_4!}(\rho_1 + \rho_2 + \rho_3 + \rho_4)^{C_4}$$
$$\times (1-\alpha_3) \sum_{i=C_4}^{C_4+G_3} (\rho_1 + \rho_2 + \rho_3 + \alpha_3\rho_4)^{C_3-C_4} + PB_3 \quad (4.18)$$

The blocking probability of class 1 comes from (4.10). Eqn. (4.16)-(4.18) are calculated from (4.11), which represents the general form of the call blocking probabilities of different traffic classes (without traffic class 1).

The arrival ratio of the different traffic classes is considered fixed such as, 1:2:4:6 throughout the analysis. The analysis of UBT scheme is experimented on this fixed call arrival rate ratio. The guard band is chosen as 90/100/110/120 which indicates that band for class 1 is 120, for class 2 is 110, and so on. So, $G_1=G_2=G_3=9$ is chosen for analysing the performances. Else, it should be mentioned to clarify the idea of considering the acceptance factors that class 4 is fractionally accepted by acceptance factor, $\alpha_3$, class 3 by $\alpha_2$, and class 2 by $\alpha_1$. There could be a number of combinations of acceptance factors to accept the calls and their results on call blocking probabilities are also different. Such experimental results are shown by Fig. 4.6, Fig. 4.7, and Fig. 4.8.

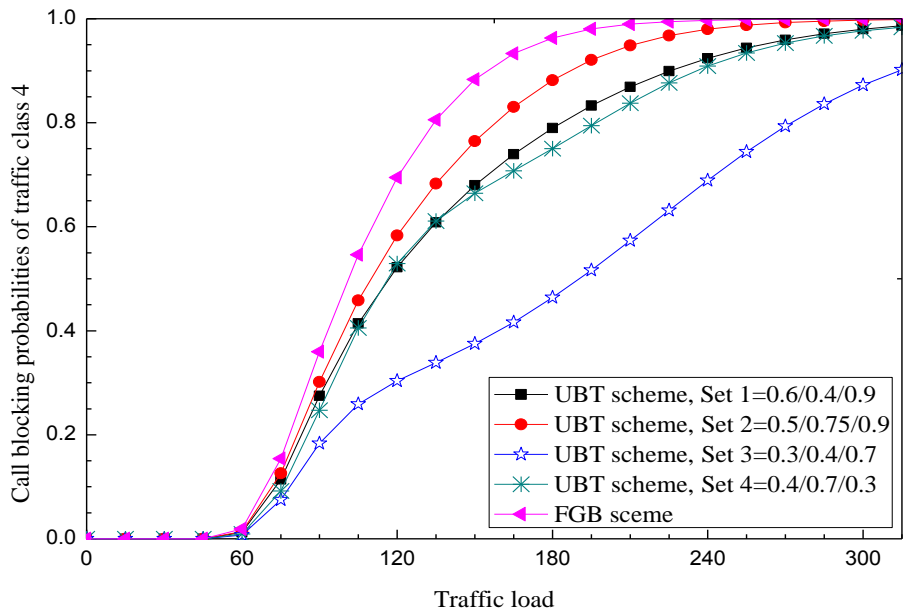

**Fig. 4.6:** Call blocking probabilities of traffic class 4 for different acceptance factors

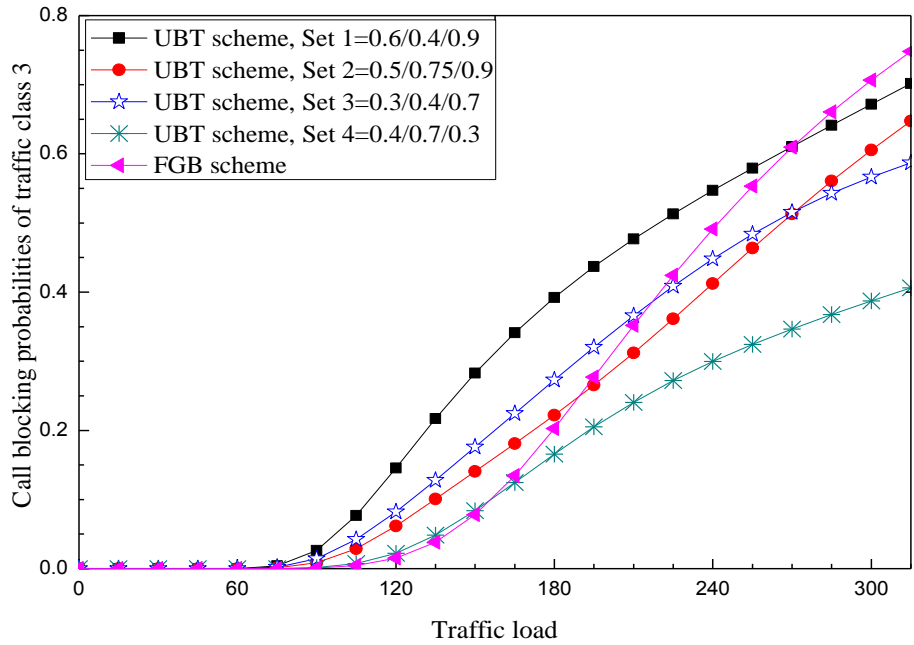

**Fig. 4.7:** Call blocking probabilities of traffic class 3 for different acceptance factors

From Fig 4.6 to Fig 4.8, it is mostly noticeable that different sets of acceptance factors represent different sets of blocking probabilities. It is commonly obeying formulation that there is trade off relations among the blocking probabilities [26] of multiple classes. The figures clarify us that if a specific set of acceptance factor decrease the blocking probability of any class, may cause of increasing the blocking probabilities of other classes. That is why, it is necessary to determine the threshold values of blocking probabilities of different classes above which the set of acceptance factors is not granted on the basis of service quality. Here, traffic class 1 does not appear because the variation of acceptance factors hardly impact on the variation of the blocking probability of traffic class 1.

In Fig. 4.9, the channel utilizing performances of UBT scheme in different acceptance factors are presented. In this figure, it is observed that the channel utilizations by the different acceptance factors of UBT scheme are more than the FGB scheme. This result depicts that if the set of acceptance factors becomes incorrect slightly the UBT scheme can do a good role in channel utilization though the QoS hampering may occur. So, in any set of acceptance factors of the UBT scheme is better that the FGB scheme in aspect of channel utilization.

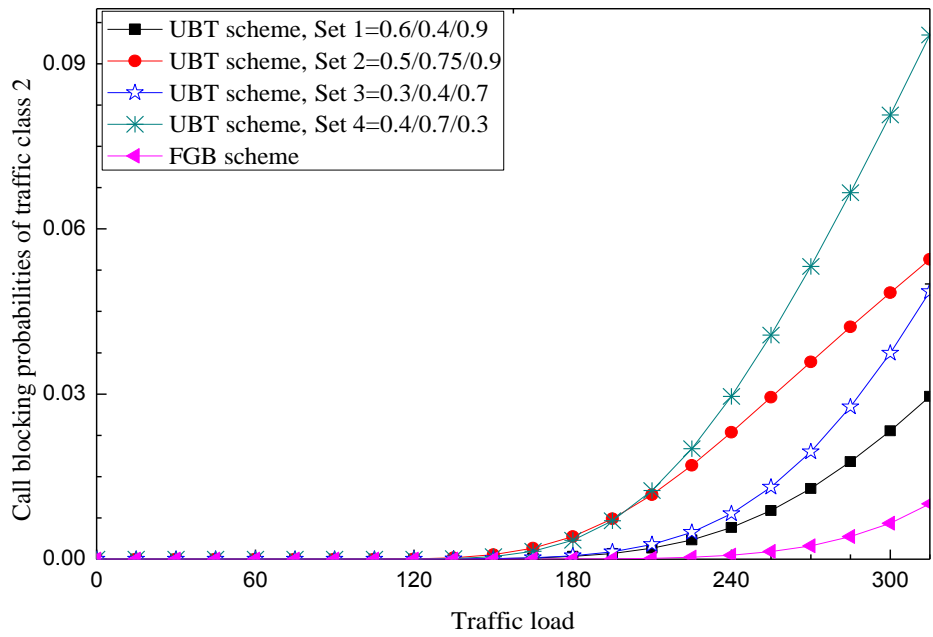

**Fig. 4.8:** Call blocking probabilities of traffic class 2 for different acceptance factors

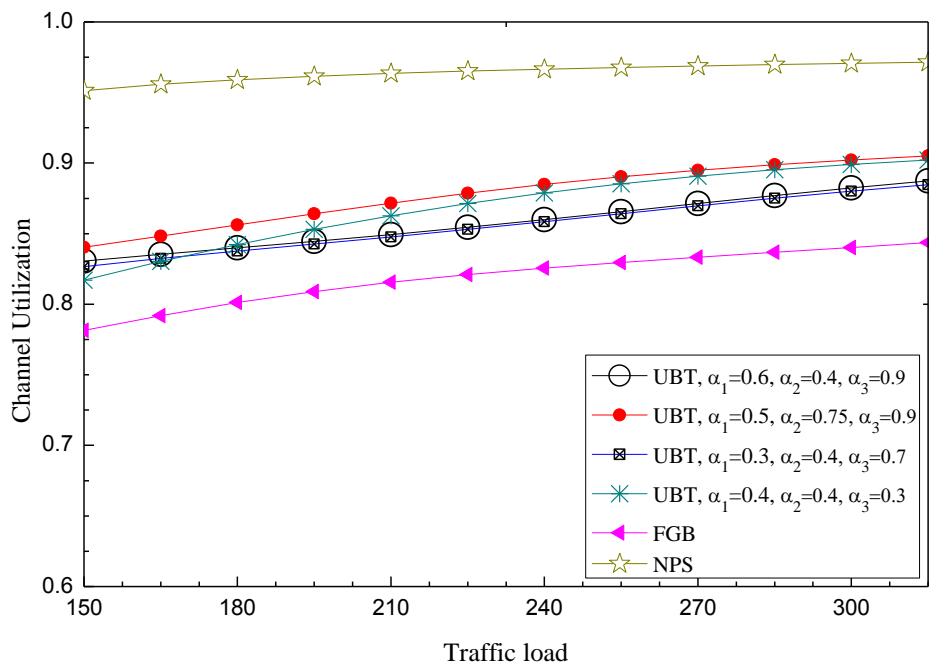

**Fig. 4.9:** Comparing the channel utilizations of UBT schemes at different sets of acceptance factors with FGB scheme and NPS

In this proposed scheme, it is necessary to find the values of acceptance factors that show the minimum call blocking probability maintaining the QoS at desired level. The call acceptance factors for different bands will be different fractional values. In this case, the values of acceptance factors, $α_1$, $α_2$, and $α_3$ are varied from 0/0/0 to 1/1/1 with possible all combinations of fractional values among them are analyzed by iterative method. By this way, the best set of acceptance factors is chosen and kept it in memory. Such a procedure is described by the block diagram in Fig. 4.4.

There are several sets of acceptance factors that execute the lower call blocking probabilities of the lower traffic classes at different pattern by maintaining the blocking probabilities of the higher traffic classes at almost ilk. Such an originated set is $α_1=0.2$, $α_2=0.3$, $α_3=0.9$ and another important set is $α_1=0.3$, $α_2=0.2$, $α_3=0.9$. In addition, it should be mentioned that the same acceptance factor for the different bands decreases the QoS and channel utilization. This set of values of acceptance factors may be different for the call arrival rate and the band is chosen for different classes. Both the call arrival rate and the channel reservation are concerning issue to get the exact set of acceptance factors.

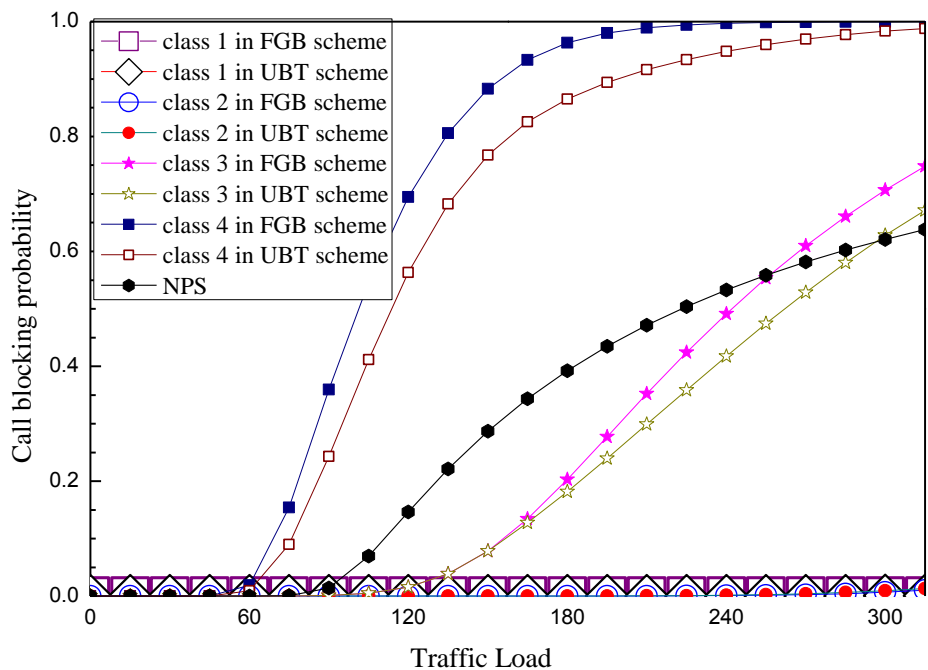

**Fig. 4.10:** Comparison of CBP of four traffic classes among NPS, FGB, and UBT scheme when $α_1=0.2$, $α_2=0.3$, and $α_3=0.9$

The above mentioned sets $α_1=0.2$, $α_2=0.3$, $α_3=0.9$ and $α_1=0.3$, $α_2=0.2$, $α_3=0.9$ are represented by Fig. 4.10 and Fig. 4.11, respectively. In Fig. 4.10, it is noticeable that, the blocking probabilities of traffic class 4 and class 3 are decreased and obviously the blocking probabilities of class 1 and class 2 remain almost constant. Fig. 4.11 shows that the blocking probability of class 3 decreases at higher traffic but slightly increases in lower traffic. Moreover, the performances of the proposed scheme are compared with NPS too.

The channel utilization by the system is focused by Fig. 4.12. From this figure, it is observed that, the channel utilization profile of the proposed UBT scheme is better than the FGB scheme. Though unused channels increase the system cost, some channels are to reserve in FGB scheme for providing the desired QoS. If proper utilization of radio resource is not done, it may be a concerning issue of rising system cost. The proposed UBT scheme increases the channel utilization without hampering the QoS. So, the UBT scheme minimizes the system cost too.

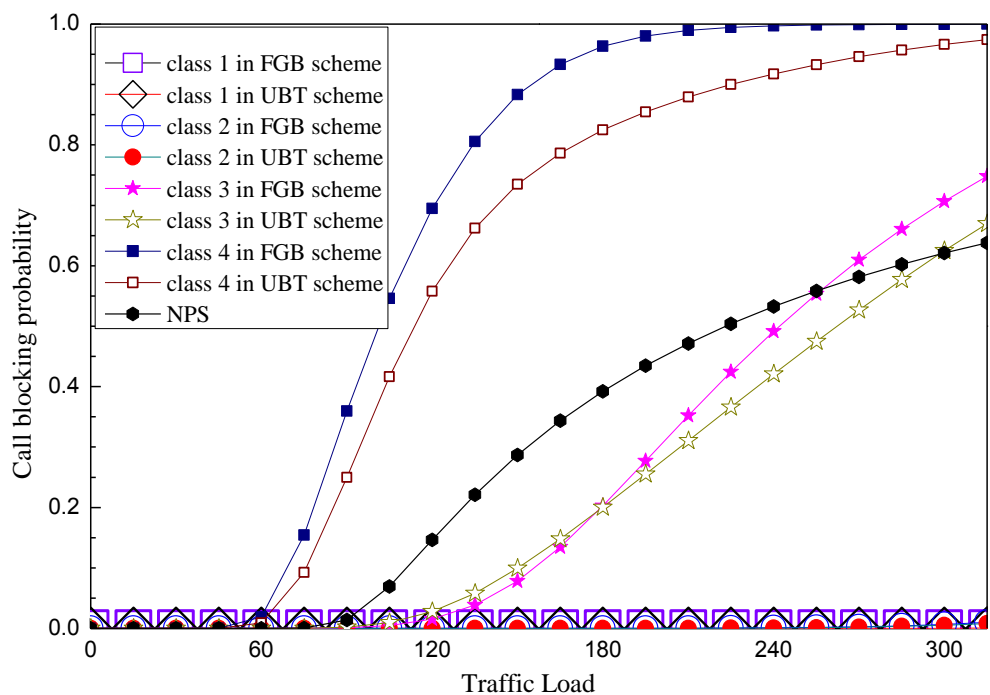

**Fig. 4.11:** Comparison of CBP of four traffic classes among NPS, FGB, and UBT scheme when $α_1=0.3$, $α_2=0.2$, and $α_3=0.9$

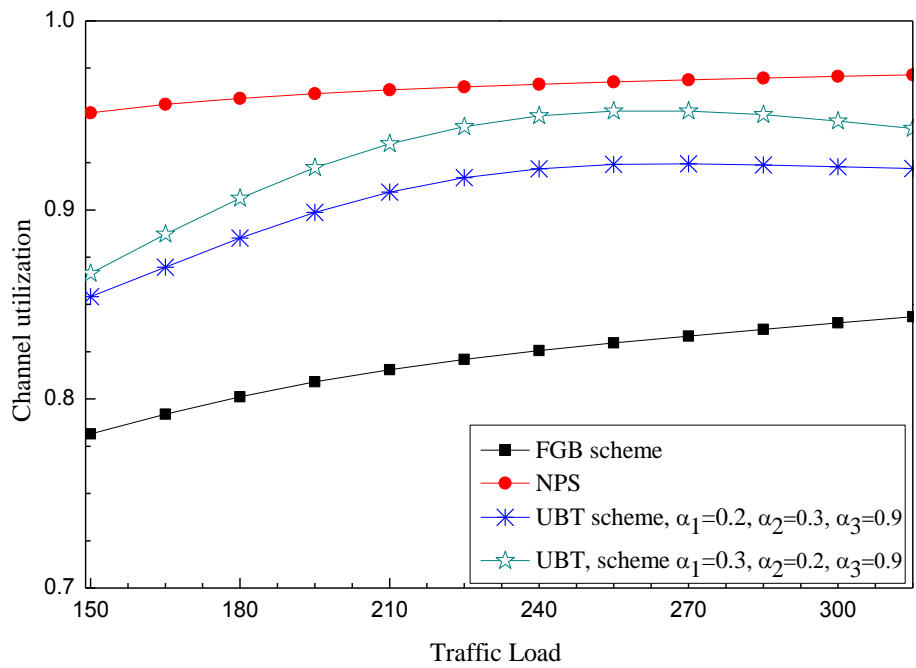

**Fig. 4.12:** Comparison of channel utilization among NPS, FGB, and UBT schemes with respect to traffic load at proper acceptance factors

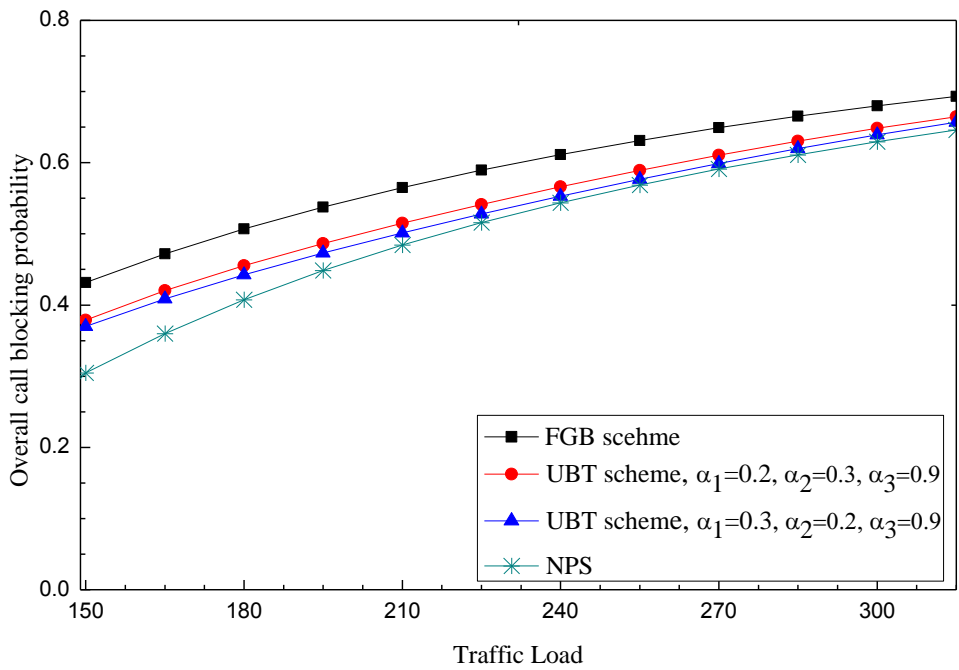

**Fig. 4.13:** Comparison of overall call blocking probabilities of NPS, UBT and FGB schemes

Inasmuch, the blocking probabilities of the class 3 and class 4 are decreased in noticeable range, the overall call blocking probability is also decreased by UBT scheme. Such a comparison is presented by the Fig. 4.13. From this figure, we detect that the overall call blocking probability of the proposed scheme is lower than the FGB scheme and at very high traffic rate the overall blocking probabilities of UBT scheme is slightly high than the non-priority scheme where the non-priority scheme shows the lowest overall call blocking probabilities. Reduced overall call blocking is another important factor in performance measurement because reduced blocking lessens the system cost [47]. Moreover, the call blocking and system cost functions are directly proportional in the aspect of economic analysis of wireless communication systems.

## 4.5  Summary

In this thesis paper, an efficient uniform band thinning CAC scheme has been proposed which hybridizes the idea of FGB scheme and uniform thinning technique for multi-class traffic. From the derived mathematical terminologies, the proposed scheme can be turned back to FGB scheme by considering zero acceptance factors. This proposed UBT scheme ensures a minimum permissible blocking probabilities of lower priority traffic calls keeping the call blocking probabilities of higher priority traffic almost same as FGB scheme. Besides, the proposed UBT scheme utilizes more channels than FGB. Consequently, overall call blocking probability is decreased by the proposed scheme in good manner with respect to FGB. From lower traffic rate to higher traffic rate the behavior of the proposed scheme maintains the same characteristics.

This work also elucidates to choose the best set of acceptance factors that improve the channel utilization and decrease the overall blocking probability as well as obviously not to hamper the QoS of higher priority calls. The process of finding the best values of acceptance factors is also described. So, this CAC scheme is undoubtedly applicable in wireless multiservice networks.

There is a suggestion to use this scheme in practice that it should be the major concerning issue that the QoS of higher priority traffic class cannot be hampered. If the system faces such a problem, it may consider the acceptance factor as zero for the consecutively nearer traffic class admission such as, if the acceptance factor of traffic class 2 becomes zero the blocking probability of traffic class 1 will be reduced. This suggestion can be placed as a special condition of the algorithm that is used to choose the set of values of acceptance factors.

# CHAPTER 5

# Conclusions

## Chapter Outlines

- ❖ Outcomes
- ❖ Discussion
- ❖ Future work

## 5.1 Thesis outcomes

Based on the simulation results, some addressable outcomes from this research can be listed as follows:

- A new guard band scheme named by UFB scheme is proposed for handover priority scheme and its mathematical expressions are explained.
- The process to reach the optimum QoS by using this UFB scheme is elucidated on the variation of uniform acceptance factor.
- The concept of calculating the fixed ratio handover call rate and the statistical rate of handover call rate are analysed, and the problems arisen due to aforesaid two different ideas are clarified.
- A new style of CAC scheme named by UBT scheme is proposed with neat mathematical expression for multiple services wireless networks.
- The mathematical expressions are presented as general form to implement the UBT scheme for any number of traffic classes.
- Optimization technique of UBT scheme is described on the basis of choosing the best set of call acceptance factors.
- The performance of UBT scheme is analysed and compared with the FGB scheme for multiservice wireless networks, and it is shown that the proposed scheme is optimum than FGB scheme.
- These two proposed schemes are designed for not only the optimum reduced call blocking probability but also the optimum utilization of radio resources.

## 5.2    Discussion

In a system, the radio resource is limited. For this reason, providing a priority to one class in its call admission is a cause to increase the call blocking probabilities of other classes. Since, handover call arrival rate is practically much more less than the new call arrival rate, a number of channels reserved for handover calls is a cause to reduce the resource utilization. In this work, some channels from the reserved channels are fractionized uniformly and the new calls are accepted by the channel with a uniform acceptance rate. In this case, this uniform rate may be the cause of increasing the handover call dropping probability which decreases the quality of services. So, it becomes a major duty to choose the band length (the amount of cells belongs to the band) and the acceptance factor. To reach this goal, it is necessary to find out the average handover call rate, the priority level of handover call rate, total number of channels belongs to the system. Generally, the band length and the value of acceptance factor vary according to the variation of the aforesaid parameters.

In case of multi-class traffic, same problem may arise. The solution is slightly difficult than the handover priority scheme. In multiple service classes oriented traffic system, the uniform call acceptance factors will be different for the different service classes at optimum conditions. The set of proper call acceptance factors may be varied according to the variation of call arrival ratio. This problem can be figured out by placing an algorithm inside the system to find the proper set of call acceptance factors.

## 5.3  Future work

This uniform fractional band scheme is not investigated in multidimensional Markov process. There is scope to analyse this idea under multidimensional Markov process and to find out the effect on performances of this scheme and its curse of dimensionality. Else, degrading rate of fractionizing band call admission strategy can be analysed for QoS provisioning in future work.

# List of Publications

*Conference Paper*

[1] **Md. Asadur Rahman**, Md. Arif Hossain, Shakil Ahmed, and Mostafa Zaman Chowdhury, "A new guard-band call admission control policy based on acceptance factor for wireless cellular network," *International Conference on Informatics, Electronics and Vision*, May 2014.

[2] **Md. Asadur Rahman**, Abu Shami Md. Zadid Shifat, and Mostafa Zaman Chowdhury, "Uniform band thinning call admission control for QoS provisioning in wireless networks," *International Conference on Computer, Information and Technology,* December 2014.

*Journal Paper*

[1] **Md. Asadur Rahman** and Mostafa Zaman Chowdhury, "Uniform fractional band CAC scheme for QoS provisioning in wireless networks," *International Journal on Innovative Computing, Information and Control*. (Under Review)